\documentclass[sigconf,noacm]{acmart}
\usepackage[ruled]{algorithm2e}
\usepackage{multirow}
\usepackage{adjustbox}

\AtBeginDocument{%
  \providecommand\BibTeX{{%
    \normalfont B\kern-0.5em{\scshape i\kern-0.25em b}\kern-0.8em\TeX}}}

\usepackage[utf8]{inputenc}
\usepackage{graphicx}
\usepackage{caption}
\usepackage{subcaption}
\usepackage{xcolor}
\usepackage{algpseudocode}
\usepackage{float}

\DeclareMathOperator{\oP}{P}

\newcommand{\ayse}[1]{\textcolor{orange}{#1}}

\newcommand{\M}{\mathcal{M}}

\DeclareMathOperator{\Range}{Range}

\DeclareMathOperator*{\argmax}{arg\,max}

\renewcommand\footnotetextcopyrightpermission[1]{}
 \makeatletter
 \def\runningfoot{\def\@runningfoot{}}
 \def\firstfoot{\def\@firstfoot{}}
 \makeatother 
 \makeatletter
 \renewcommand\@formatdoi[1]{\ignorespaces}
 \makeatother

\copyrightyear{}
\acmYear{2023}
\acmConference[ACM CCS]{ACM CCS}{2023}{Denmark}
\acmBooktitle{Anonymous Conference}
\acmDOI{}
\acmISBN{}

\begin{document}

\title{Node Injection Link Stealing Attack}

\author{Oualid Zari}
\email{oualid.zari@eurecom.fr}
\affiliation{%
  \institution{Eurecom}
  \city{Biot}
  \country{France}
}

\author{Javier Parra-Arnau}
\email{javi.parra-arnau@kit.edu}

\affiliation{%
  \institution{Karlsruhe Institute of Technology}
  \city{Karlsruhe}
  \country{Germany}
}
\affiliation{%
  \institution{Universitat Politecnica de Catalunya}
  \city{Barcelona}
  \country{Spain}
}

\author{Ayşe Ünsal}
\email{ayse.unsal@eurecom.fr}
\affiliation{%
  \institution{Eurecom}
  \city{Biot}
  \country{France}
}
\author{Melek Önen}
\email{melek.onen@eurecom.fr}
\affiliation{%
  \institution{EURECOM}
  \city{Sophia-Antipolis}
  \country{France}
}



\begin{abstract}
In this paper, we present a stealthy and effective attack that exposes privacy vulnerabilities in Graph Neural Networks (GNNs) by inferring private links within graph-structured data. Focusing on the inductive setting where new nodes join the graph and an API is used to query predictions, we investigate the potential leakage of private edge information. We also propose methods to preserve privacy while maintaining model utility. Our attack demonstrates superior performance in inferring the links compared to the state of the art. Furthermore, we examine the application of differential privacy (DP) mechanisms to mitigate the impact of our proposed attack, we analyze the trade-off between privacy preservation and model utility. Our work highlights the privacy vulnerabilities inherent in GNNs, underscoring the importance of developing robust privacy-preserving mechanisms for their application.
\end{abstract}







\maketitle
\pagestyle{empty}

{\section{Introduction}

Graph-structured data has become increasingly prevalent in today's data-driven landscape, particularly in applications involving social networks, biological systems, or recommendation engines. Graph Neural Networks (GNNs) have emerged as powerful tools for analyzing and learning from such data, offering remarkable performance in various tasks. These advantages both in efficiency and utility unfortunately come with a high cost in terms of privacy since the underlying graph structure is usually considered as sensitive information. For example, in a social network, the link between nodes usually represents whether users share common interests, beliefs, political views or even sexual preferences. This unfortunately leaks important information about users and can cause serious damage to their privacy \cite{camb_analytica}.
In this paper, our primary objective is to advance the understanding of edge privacy in GNNs by first developing a novel link-stealing attack, named Node Injection Link Stealing (NILS) attack, and then proposing a tailored Differential Privacy notion to protect against this attack. In this context, the adversary aims to infer the links among a set of target nodes. We focus on a specific scenario: training the GNN model for node classification tasks. In this scenario, the GNN processes input data consisting of the graph structure and node features, subsequently producing predictions that indicate the class membership of each node. This model will be served at inference time through an inference API.

Early works, such as the Linkteller attack \cite{linkteller}, demonstrated that by probing the features of nodes and analyzing the output predictions generated by the GNN, an attacker could successfully infer the links of the graph. Another study in ~\cite{he2021stealing} focused on the correlation of nodes' features in order to infer the links between them.
Unlike those previous works, we propose a stronger adversary who takes advantage of the dynamic nature of GNNs. The adversary adds a new node to the graph, connects it to the target node through a single edge/link} and further queries the model with some malicious input features that are generated following different strategies. Establishing this new connection allows the attacker to infer some of the target node's neighbors and hence steal a subset of the graph's connections. The various adversary strategies mainly differ with respect to the actual values of the input feature vectors.

Similar to previous work \cite{linkteller}, we further study potential defense strategies mainly based on differential privacy (DP). More specifically, we first propose a dedicated privacy notion specifically crafted to counter the proposed adversary, and evaluate existing DP-based defense strategies under this privacy definition, which we call \textbf{one-node-one-edge privacy}.

To summarize, we make the following contributions:
{\begin{itemize}
  \item We propose a novel attack NILS for inferring private links in a graph structure by injecting a new node, linking it to a target node, and employing various attack strategies to analyze the changes in the GNN's output;
  \item We provide a comprehensive evaluation of the proposed attack's effectiveness on various datasets, demonstrating its superior performance compared to existing work such as LinkTeller \cite{linkteller} and link-stealing \cite{he2021stealing};
  \item We explore the application of DP mechanisms as a means to mitigate the effectiveness of our proposed attack, evaluating the trade-off between privacy preservation and model utility. To this end, we introduce a new notion of privacy and evaluate defense strategies under this new notion.
\end{itemize}

\section{Background\label{sec:background}}
In this section, we give a brief introduction to GNNs, and formalize the concept of DP.
\subsection{Graph neural networks}

\subsubsection{GNNs Overview}
\label{sec:GNNsOverview}

GNNs \cite{gnn_first} have emerged as a powerful class of machine learning models specifically designed to handle graph-structured data. They have gained considerable attention due to their ability to effectively learn and capture complex patterns in graph data, showing significant performance across a wide range of tasks, such as node classification \cite{Petar_GTA_node_class, Hamilton_node_class}, link prediction \cite{Zhang_link_pred}, and graph classification \cite{Xu_graph_class, Wang_graph_class}. In this paper, we specifically focus on the task of node classification, where the objective is to assign labels to individual nodes based on their features and on the overall graph structure.

More specifically, a graph $G = (V, E)$  is defined as a collection of nodes $V$ and edges $E$. Nodes represent data points such as users in social networks or proteins in biological networks, while edges represent relationships or interactions between the nodes. Graphs can be represented using an adjacency matrix $A \in \mathbb{R}^{n \times n}$, where $n = |V|$ is the number of nodes in the graph, and $A_{ij} = 1$ if there exists an edge between nodes $i$ and $j$, and $A_{ij} = 0$ otherwise.
Additionally, nodes exhibit a set of features, which can be represented as vectors containing $d$ elements, where $d$ corresponds to the number of features. In social networks, these features may include demographic information such as age, gender, and location, as well as user interests and preferences. To capture these features, the graph is also associated with a feature matrix $X \in \mathbb{R}^{n \times d}$. This matrix provides essential information about the characteristics of each node in the graph.

GNNs primarily operate by employing a message-passing mechanism \cite{gnn_first} that allows nodes to exchange and aggregate information from their local neighborhoods. This iterative process helps GNNs capture local and global structural information in the graph. For instance, in the context of graph convolutional networks (GCNs) \cite{gcn_first}, the most representative and well-established GNNs models, their core architecture consists of a series of graph convolutional layers, which can be formulated as follows:
\begin{equation}
H^{(0)}= X, \ayse{\quad}
H^{(l+1)} = \sigma \left( \hat{A} H^{(l)} W^{(l)} \right), \ayse{\quad}
H^{(L)} = P
\end{equation} 
For instance, $H^{(0)}$ denotes the node feature matrix $X$; $H^{(l)} \in \mathbb{R}^{n \times d_l}$ is the hidden node representation matrix at layer $l$, where $L$ is the total number of layers; and $P\in\mathbb{R}^{n \times\ c}$ represents the prediction scores for each potential class or label associated with the queried nodes, where $c$ reprensents the number of classes; $W^{(l)} \in \mathbb{R}^{d_l \times d_{l+1}}$ is the learnable weight matrix for layer $l$; $\sigma(\cdot)$ is an activation function (e.g., ReLU), and $\hat{A}$ is a normalized adjacency matrix.

\subsubsection{GNNs with dynamic graphs}
GNNs usually handle dynamic graph data as in real-life scenarios such as social network applications or recommendation systems, where graphs usually evolve over time. New nodes or edges may be introduced with time and the goal would be to make predictions for such new nodes.

When a new node is added to the graph, both the adjacency matrix  $A \in \mathbb{R}^{n \times n}$, and the feature matrix $X \in \mathbb{R}^{n \times d}$ are updated. The adjacency matrix expands to $A' \in \mathbb{R}^{(n+1) \times (n+1)}$, while the feature matrix becomes $X' \in \mathbb{R}^{(n+1) \times d}$,  incorporating the new node's connections and features, respectively.

Once the graph is updated, the GNN performs inference on the modified graph, using the message-passing mechanism described earlier.
\begin{table}[h]
\centering
\begin{tabular}{ll}
\toprule
Symbol & Description \\
\midrule
$A$ & Adjacency matrix \\
$\mathcal{A}$ & Adversary \\
$E$ & Set of edges in the graph \\
$G$ & Graph \\
$n$ & Number of nodes \\
$V$ & Set of nodes in the graph \\
$V_\mathcal{A}$ & Target set nodes \\
$v_m$ & Malicious injected node \\
$v_t$ & Target node \\
$P$ & Prediction scores of the GNN \\
$X$ & Feature matrix \\
$x_t$ & Features of the target node \\
$x_m$ & Features of the malicious node \\
\bottomrule
\end{tabular}
\caption{List of notations.}
\label{tab:notations}
\end{table}

\subsection{Differential privacy}
\label{sec:background:DP}
The original definition of DP~\cite{Dwork06A,dwork2014algorithmic} was introduced in the context of microdata, that is, databases containing records at the level of individuals. A central aspect of DP is the concept of \textit{neighborhood}, which was defined originally for that data structure as follows. 
\begin{definition}[\textbf{Neighboring databases}] 
Let $\mathcal{D}$ be the class of possible databases. Any two databases $D,D'\in \mathcal{D}$ that differ in one record are called \emph{neighbors}. For two neighbor databases, the following equality holds: $$d(D,D') = 1,$$ where $d$ denotes the Hamming distance. 
\end{definition}

\begin{definition}[\textbf{$(\varepsilon, \delta)$-Differential privacy} \cite{Dwork06A,dwork2014algorithmic}]
	\label{def:dp}
	A randomized mechanism $\mathcal{M}$ satisfies $(\varepsilon,\delta)$-DP with $\varepsilon,\delta \geqslant 0$ if, for all pairs of neighboring databases $D,D'\in \mathcal{D}$ and for all measurable $\mathcal{O}\subseteq \Range(\mathcal{M})$,
	
	$$\oP\{\mathcal{M}(D)\in \mathcal{O}\} \leqslant e^{\varepsilon} \oP\{\mathcal{M}(D')\in \mathcal{O}\} + \delta.$$
\end{definition}

In words, the output of a mechanism satisfying DP should not reveal the presence or absence of any specific record in the database, up to an exponential factor of $\varepsilon$. When each record corresponds to a distinct individual respondent, DP aims to ensure their information will remain confidential. A lower value of $\varepsilon$, referred to as the \emph{privacy budget}, provides stronger protection.

Probably the most popular mechanism satisfying DP is the Laplace mechanism, which relies on a quantity called \emph{global sensitivity}, defined next.
\begin{definition}[$L_p{\text -}$\textbf{Global sensitivity}~\cite{dwork2014algorithmic}]
\label{def:GS}
The $L_p$-global sensitivity of a query function $f\colon\mathcal{D} \rightarrow \mathbb{R}^d$ is defined as
\begin{equation*}
\Delta_p(f)=\max_{\substack{\forall D,D'}\in\mathcal{D}} \| f(D) - f(D')\|_p,
\end{equation*}
where $D,D'$ are any two neighbor databases.
\end{definition}

\begin{definition}[\textbf{Laplace mechanism}~\cite{dwork2014algorithmic}] Given any function $f\colon\mathcal{D} \rightarrow \mathbb{R}^d$, the Laplace mechanism mechanism is defined as follows:
$$\mathcal{M}_L(D,f(\cdot),\varepsilon) =f(D) + (Y_1,\ldots,Y_d),$$
where $Y_i$ are i.i.d. random variables drawn from a Laplace distribution with zero mean and scale
$\Delta_1(f)/\varepsilon$.
\end{definition}

\section{Related work\label{sec:related_work}}

GNNs have gained significant attention in recent years due to their effectiveness in handling graph-based data across various applications \cite{gnn1, gnn_first, gnn2, gnn3}. As the adoption of GNNs increases, concerns regarding privacy and adversarial attacks on these networks also arise and become more significant \cite{sun2018adversarial_survey, dai2022comprehensive_survey}. On the other hand, several privacy-preserving methods are developed to mitigate the effectiveness of these privacy attacks against GNNs \cite{DP_GAP, sok_DP}.
\subsection{Privacy attacks on GNNs}
Privacy attacks on GNNs can be categorized based on the actual leakage in the graph, namely, information about graph nodes, their attributes, or graph edges. Node privacy attacks, such as membership inference attacks (MIA) \cite{MIA_GNN, wu2021MIA, conti2022label_MIA, olatunji2021MIA}, aim to determine if a given node was part of the training set. In contrast, attribute inference attacks \cite{linkstealing_reconstruction} focus on revealing sensitive information related to node attributes, violating attribute privacy. In this work, we concentrate on edge privacy violation, where the common attacks are the so-called link stealing, re-identification, or inference attacks, which aim to uncover the edges of the graph structure used by the GNN.

Early works \cite{he2021stealing, linkstealing_reconstruction, linkteller} have demonstrated the success and feasibility of link-stealing attacks. In the attack proposed in \cite{he2021stealing}, the adversary leverages prior knowledge about the graph, such as the likelihood of nodes with similar features or predictions being connected, to infer links in the graph. The attacker applies methods such as clustering to predict connections for nodes within the same cluster. In \cite{linkstealing_reconstruction}, the authors demonstrate that by accessing the node embeddings trained to preserve the graph structure, one can recover edges by training a decoder to convert the embedding to the graph structure. The Linkteller attack \cite{linkteller} involves probing the features of the nodes and studying their output predictions by the GNN to infer the links of the graph.

Existing link-stealing attacks exhibit certain weaknesses. The attack in \cite{he2021stealing} assumes a powerful adversary who requires access to the features, a shadow dataset, and the ability to train shadow GNNs in order to train an attack model. The attack model is trained to classify the link presence based on the output predictions or features. The attack's performance declines in the inductive setting, where training and inference occur on different graphs, as evidenced in the further Linkteller paper \cite{linkteller}. Additionally, its effectiveness diminishes when there is no correlation between features and links of the nodes. 

On the other hand, the main drawback of the Linkteller attack \cite{linkteller} is its non-stealthy perturbation of features, particularly when dealing with discrete datasets. The Linkteller's strategy consists of altering the input features of the graph to obtain information about the links. For discrete datasets, the perturbation can render the features as real values, making them easier to be detected. Moreover, the effectiveness of the Linkteller attack decreases when mounted against deep GNNs with depths of more than three.

In addition to privacy attacks targeting GNNs, adversarial attacks exist where the adversary's goal is to deceive the GNN's predictions or degrade its utility. These attacks involve altering the graph structure through node addition or deletion \cite{zhang2020_adversarial, mu2021hard_adversarial}.

In this paper, we propose a novel link-stealing attack NILS that addresses the limitations of existing approaches, taking advantage of the dynamic nature of GNNs by injecting malicious nodes in the style of an adversarial attack. Our proposed NILS attack outperforms previous link-stealing attacks \cite{he2021stealing, linkteller}. 

\subsection{Differential privacy mechanisms for graphs}

DP has been extensively studied and applied to various data types, including graphs, with the aim of preserving sensitive information. Various DP mechanisms have been developed \cite{node_dp_Brunet, node_dp_Daigavane} to protect both node and edge information. Node-level DP focuses on preserving the privacy of individual nodes, protecting from attacks, such as membership inference attacks \cite{MIA_GNN, wu2021MIA, conti2022label_MIA}. In contrast, Edge-level DP seeks to preserve the privacy of edge information, which represents relationships between nodes, preventing link stealing attacks \cite{linkstealing_reconstruction, linkteller, he2021stealing}.

Substantial research has been conducted on achieving node-level DP and edge-level DP in graph-based models. Several approaches allow for the publication of graph statistics with edge-level DP guarantees, including degree subgraph count \cite{Karwa_subgraph_count}, and degree distributions \cite{Hay_degree_dist, WeiYen_degree_dist}. Although these statistics are beneficial for graph analysis, they are inadequate for training a GNN model, as most of the GNNs require access to the raw graph structure for the message-passing mechanism. Consequently, other approaches have been developed to train GNN models by adopting input perturbation DP, releasing the graph while ensuring edge-level DP \cite{linkteller, Rui_adj_release, Hiep_adj_release}

Furthermore, when designing DP solutions, it is crucial to consider specific privacy threats and adversary strengths. In the context of our proposed NILS attack, the adversary is capable of injecting nodes into the graph to discover sensitive edge information, violating edge privacy. Therefore, we propose a customized DP notion that specifically addresses this type of privacy attack. We then leverage the LapGraph algorithm \cite{linkteller} to achieve the desired DP guarantees under the new, tailored notion and study its effectiveness.

\section{Node Injection Link Stealing Attack\label{sec:attack}}

GNNs are prone to various privacy attacks that usually aim at learning as much information as possible about their underlying graph structure. GNNs inherit the potential attacks against standard neural networks such as membership inference attacks \cite{MIA_GNN, MIA_Jiayuan}, whereby the goal of the adversary is to ascertain whether a sample is included in the training dataset or not.

As introduced earlier, in this paper, we focus on a particular attack named as \textit{link stealing attack}, where an adversary without access to the adjacency matrix aims to learn whether a particular edge exists or not.

In this section, we first introduce the threat model to characterize the adversary's background knowledge. Then, we propose our node injection link stealing attack that takes advantage of the dynamicity of GNNs.

\subsection{Threat model}
\subsubsection{Environment}
As mentioned in the previous section, we consider a GNN application in which a server has already trained the GNN using a specific dataset and offers access to this GNN through a black-box API. In this context, the black-box API is an interface provided by the server that enables users to interact with the pre-trained GNN model without directly accessing its internal components, such as the model architecture, parameters, or graph structure.
Users can submit prediction queries using node IDs. If a new node needs to be added to the graph, users can employ a \textit{connect} query to attach the node to the graph before querying its prediction based on its ID.
The API processes input data into output predictions, ensuring that the model's underlying computations remain hidden from the user. Users can query this GNN for the purpose of node classification. Hence, the query consists of the queried node's ID and the output of this query is the vector of prediction scores for this particular node. The users do not have the knowledge of edges of this graph. Hence the only information that a user knows is the set of nodes' ids.

\subsubsection{Adversary's goal and knowledge}

We consider an adversary, $\mathcal{A}$, who assumes the role of a GNN user. Her objective is to determine the neighbors of a specific \textit{target node}, $v_t$, selected from a set of \textit{target nodes}, $V_{\mathcal{A}}$, within the graph. This is done based on the GNN's predictions for the node set $V_{\mathcal{A}}$. In simpler terms, $\mathcal{A}$ aims to identify the neighbors of the target node $v_t$ that are included in the target set nodes $V_{\mathcal{A}}$.

We should note that if the adversary aims to identify all the links within the graph, then the set of target nodes $V_{\mathcal{A}}$ becomes the set containing all the nodes of the graph $V$. To achieve this, the adversary may need to perform multiple node injections, targeting different nodes from the graph each time. However, the practicality of such an approach is debatable. The adversary's selection of target nodes reflects her background knowledge about these nodes. For instance, in the context of social networks, the adversary's background knowledge could include information such as users' interests. This information can guide the adversary in selecting target nodes $V_{\mathcal{A}}$ that are more likely to be connected. In our attack scenario, we choose the target nodes uniformly at random.

The adversary $\mathcal{A}$ is able to obtain the predictions of the target nodes $V_{\mathcal{A}}$ by sending the server their corresponding IDs through the provided API.
In addition, the adversary $\mathcal{A}$ is able to use the \textit{connect} query to connect a node $v_m$ to a target node $v_t$. In general, we assume that the adversary does not have access to the features of the nodes in the graph, with the exception of certain attack strategies described in Sec.~\ref{subsection:malicious features strategies}.

\subsection{Node injection link stealing attack}
\label{sec:node-injection-attack}
In this section, we formally define our NILS attack that, unlike existing link-stealing attacks, exploits the dynamic nature of the underlying GNN. Indeed, adversary $\mathcal{A}$ can \textit{connect} new nodes and further query the prediction scores of a set of nodes $V_{\mathcal{A}}$ in the graph. While adding this new node $v_m$, $\mathcal{A}$ can choose which existing node $v_t$ it actually connects to and hence try to discover its neighbors. More formally:

\begin{enumerate}
\item $\mathcal{A}$ first queries the prediction scores of the target nodes $V_{\mathcal{A}}$
and receives the corresponding prediction matrix $P$ of the target nodes $V_{\mathcal{A}}$.
    \item $\mathcal{A}$ generates malicious features of a malicious node $v_m$ based on the obtained prediction matrix $P$ (see Sec.~\ref{sec:strategies} for further details on this step).
    \item Next, $\mathcal{A}$ sends a \emph{connect} query to inject the malicious node $v_m$. The query has the following parameters: the features $x_m$ of the new node, and the ID of the target node $v_t$ the adversary wishes to connect $v_m$ to.
    \item The server adds this \text{malicious} node $v_m$ to the graph and links it to the target node $v_t$.
    \item $\mathcal{A}$ queries back the server for new prediction matrix $P'$ of the target nodes $V_{\mathcal{A}}$ and obtains it.
    \item With access to $P$ and $P'$, $\mathcal{A}$ computes the $L_1$ distance between $P(v)$ and $P'(v)$ of each node $v$ in $V_{\mathcal{A}}$.
    A significant change in the prediction scores of a node $v$ indicates a high probability of being a neighbor to $v_t$. If the difference exceeds a threshold $R$, the adversary infers that node $v$ is a neighbor of $v_t$.
\end{enumerate}

The decision threshold $R$ is determined through an extensive parameter tuning process, aiming for an optimal trade-off between precision and recall in identifying the true neighbors of the target node. This balance is represented by the $F_1$ score. We evaluate various candidate values of $R$, selecting the one that yields the highest $F_1$ score as the optimal threshold. The results reported in our study are based on this optimal value of $R$.

 This attack strategy is depicted in Figure \ref{fig:attack_strategy} and outlined in Algorithm \ref{alg:prob_vecs_attack}.
 
\begin{figure}[h]
\centering
\includegraphics[scale=0.75, width=\columnwidth]{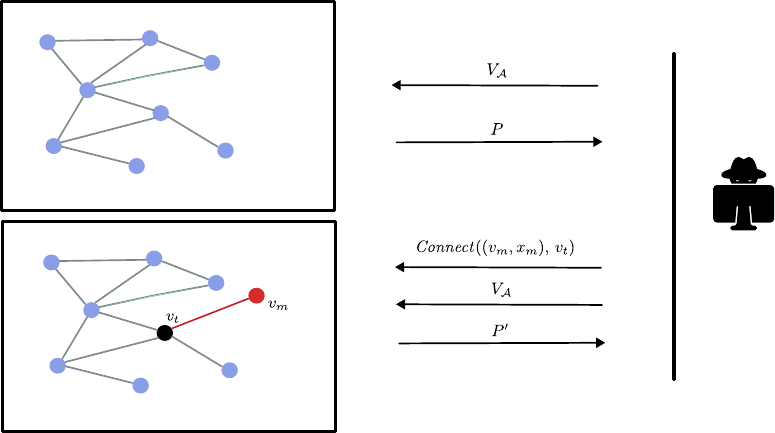}
\caption{Adversary-Server Interaction: In the inference phase, the adversary first queries the prediction scores $P$ of the target nodes, represented as $V_{\mathcal{A}}$. Next, the server sends the predictions $P$ of the GNN to the adversary. Then, the adversary sends a \textit{Connect} query to inject the malicious node $v_m$, with features $x_m$, to the target node $v_t$. Finally, after the injection, the adversary queries again the prediction scores $P'$ of the target nodes $V_{\mathcal{A}}$.}
\label{fig:attack_strategy}
\end{figure}

\begin{algorithm}
\caption{Node Injection Link Stealing Attack}
\label{alg:prob_vecs_attack}
\textbf{Input:} set of nodes $V_{\mathcal{A}}$ and target node $v_t$. \\
\textbf{Output:} the identified neighbors of $v_t$ by the adversary.\\

$P$ = GNN($V_{\mathcal{A}}$, $X_{V_{\mathcal{A}}}$) \Comment{Step 1}\\
Generate malicious features $x_m$ of node $v_m$ \Comment{Step 2}\\
Connect node $v_m$ to $v_t$. \Comment{Step 3-4}\\
$P'$ = GNN($V_{\mathcal{A}} \cup v_m$ , $X_{V_{\mathcal{A}}} \cup x_m$) \Comment{Step 5}\\
\For{each node $v$ in $V_{\mathcal{A}}$}{ 
$D(v)$ = $\lVert P(v) - P'(v) \rVert_1$ \Comment{Step 6} \\ 
\If{$D(v) \geq R$}{
      $v$ is a neighbor of $v_t$ \\}
\Else{
      $v$ is not a neighbor of $v_t$ \\}
}
\end{algorithm}

 \subsection{Strategies for malicious node's features\label{subsection:malicious features strategies}}
 \label{sec:strategies}
In order to evaluate how the injection of the malicious node $v_m$ influences the predictions of the GNN, we study five strategies to generate the malicious node's features $x_m$. This helps us to assess the success of our attack. These five strategies are designed with varying degrees of sparsity and stealthiness,
enabling us to explore their effectiveness in altering the model's predictions. We define the proposed strategies as follow:

\begin{enumerate}
    \item \textbf{All-ones strategy}: Generates a dense feature vector for the malicious node, containing all ones, as shown in the equation below:
    \begin{equation*}
        x_m = \mathbf{1}.
    \end{equation*}
    This strategy potentially causes significant changes in predictions but may be less stealthy due to its dense feature vector.
    
    \item \textbf{All-zeros strategy}: Creates a sparse feature vector for the malicious node, containing all zeros, as shown in the equation below:
    \begin{equation*}
        x_m = \mathbf{0}.
    \end{equation*}
    This approach may subtly alter the output of the GNN, leading to smaller changes in predictions, while offering increased stealthiness.
    
    \item \textbf{Identity strategy}: Introduces a malicious node with a feature vector identical to the target node's feature vector, as shown below:
    \begin{equation*}
        x_m = x_t.
    \end{equation*}
    This strategy causes confusion in the model's predictions for neighboring nodes and has variable stealthiness based on the similarity between injected and target nodes.
    For this strategy, we assume that $\mathcal{A}$ knows the features of the target node $x_t$.
    
    \item \textbf{Max attributes strategy}: This method creates a malicious node feature vector by computing the element-wise maximum of each attribute in the target nodes' feature matrix.
    Specifically, it considers only nodes from classes different from the target node's class, as shown below:
    \begin{equation*}
    x_{m,k} = \max_{i \in V_{\mathcal{A}}, \text{ with } C_i \neq C_t} X_{i,k}, \quad \text{for} \quad k = 1, \ldots, d.
    \end{equation*}
    Here, $C_i$ represents the class of node $i$, and $C_t$ is the class of the target node.
    This strategy potentially causes significant changes in predictions but may be less stealthy due to exaggerated features. We assume in this strategy that the adversary has access to the features of the set of target nodes $V_{\mathcal{A}}$ and also to their predicted classes by the GNN. The predicted classes are accessible to the adversary after step 1 in Algorithm \ref{alg:prob_vecs_attack}.
    
    \item \textbf{Class representative strategy}: This approach generates a malicious node feature vector by selecting the feature vector of the node with the highest confidence score for a specific class, different from the target node's class, as shown below:
    \begin{equation*}
    x_m = x_{i^*} \text{ with } i^* = \argmax_{\substack{ i \in V_{\mathcal{A}}, \\ C_i \neq C_t}} p_{i,j}.
    \end{equation*}
    In this equation, $x_m$ is the malicious node feature vector, $i^*$ is the node index with the highest confidence score for a specific class different from the target node's class, $V_{\mathcal{A}}$ is the set of target nodes, $C_i$ represents the class of node $i$, and $C_t$ is the class of the target node. This strategy leverages the model's predictions to alter the neighbors of the target node predictions, potentially offering increased stealthiness.
\end{enumerate}
Additionally, we introduce the \textit{so-called} LinkTeller \textbf{Influence} strategy as an alternative to the original method in \cite{linkteller} incorporating their feature perturbation strategy.
This strategy entails perturbing the features of the target node by adding a small real value $\delta$, as shown below:
\begin{equation*}
x_m = x_t + \mathbf{\delta}.
\end{equation*}
We assess the performance of the Influence strategy in comparison to other strategies, aiming to determine whether the attack performance gains are attributable to node injection or the crafting of malicious features. It is worth noting, however, that the Influence strategy may be easily detected if the feature $x_t$ has a discrete nature, given that $x_m$ is real-valued.

\section{Evaluation of The Attack\label{sec:attack_eval}}
In this section, we present the evaluation results of our proposed attack. First, we introduce our experimental setup. Then, we provide a detailed analysis of the performance of our attack on various datasets, discussing its effectiveness and limitations.
\subsection{Experimental setup}
\subsubsection{Datasets}
In order to evaluate the effectiveness of our attack, we conducted experiments on various real-world datasets previously utilized in related research. We include the Flickr \cite{ZengZSKP20_Flickr} dataset, where nodes represent images uploaded to the Flickr platform. Edges connect nodes if the images share common properties like geographic location, gallery, or user comments. Node features contain word representations. Additionally, we utilize two Twitch datasets (TWITCH-FR and TWITCH-RU)\cite{rozemberczki2021twitch} to evaluate NILS. We use Twitch-ES to train the GNNs as done previously in \cite{linkteller} for the inductive setting. Twitch datasets \cite{rozemberczki2021twitch} illustrate follow relationships between users on the Twitch streaming platform. The objective of these datasets is to perform binary classification to determine if a streamer uses explicit language, using features such as users' preferred games, location, and streaming habits.

Furthermore, for the transductive setting, where the training and testing of the GNNs occur on the same graph, we incorporate three citation network datasets \cite{citation_datasets}, Cora, Citeseer, and Pubmed. These datasets capture citation relationships among scientific publications across various fields. The classification task of these datasets involves predicting the topic of publications based on their textual features. While Cora and Citeseer encompass general scientific publications, Pubmed is dedicated to biomedical publications. By employing these datasets in our evaluation, we aim to demonstrate the effectiveness of our proposed attack in both inductive and transductive settings, as well as across a range application domains.

\subsubsection{Models}

In our study, we follow LinkTeller's approach to training the models and selecting hyperparameters \cite{linkteller}. In LinkTeller \cite{linkteller}, the authors trained Graph Convolutional Networks (GCNs) using various configurations and hyperparameters, which encompassed normalization techniques applied to the adjacency matrix, the number of hidden layers, input and output units, and dropout rates. In order to identify the optimal set of hyperparameters, the authors employed a grid search strategy, systematically exploring combinations of hyperparameters and evaluating their performance on a validation set.
The search space for hyperparameters and the formulae for different normalization techniques were provided in \cite[Appendix F]{linkteller}. After obtaining the best set of hyperparameters, the authors trained the GCN models to minimize the cross-entropy loss for the intended tasks.

In our experiments, we adhere to the same methodology as in LinkTeller \cite{linkteller}, ensuring consistency across the studies. By utilizing the same training procedures and hyperparameter tuning strategies, we aim to provide a comprehensive understanding of the attack performance across different layer configurations (two, three, and four layers) while maintaining consistency.

\subsubsection{Evaluation of attack performance}
In accordance with the evaluation methodology presented in the LinkTeller paper \cite{linkteller}, we employ precision, recall, and the $F_1$ score as our primary evaluation metrics. These metrics are particularly suitable for addressing the imbalanced binary classification problem at hand, in which the minority class (i.e., connected nodes) is of central interest. We primarily select the set target nodes $V_{\mathcal{A}}$, such that  $|V_{\mathcal{A}}|=500$, using a uniform random sampling approach. Furthermore, following the baseline \cite{linkteller} study's example, we explore scenarios where target nodes exhibit either low or high degrees. A comprehensive discussion of the sampling strategy can be found in \cite[Section V.D.]{linkteller}. We report the results averaged over three runs with different random seeds along with the standard deviation.
\subsection{Analysis of strategies for malicious node’s features}
In this section, we analyze the impact of different strategies, as defined in Section \ref{subsection:malicious features strategies}, for generating the features $x_m$ of the malicious node $v_m$ on the success of our attack.

The success rates of these strategies, as shown in Table \ref{tab:adv_strategies}, reveal that the All-ones, Max attributes, and Class representative strategies are the most effective in causing significant changes in the predictions of the target node's neighbors. These results suggest that injecting nodes with high-valued or class-specific features can effectively disrupt the model's output predictions.

Conversely, the All-zeros, and Identity strategies exhibit relatively lower success rates, as shown in Table \ref{tab:adv_strategies}. While these strategies offer certain benefits in terms of stealthiness, their impact on the graph structure and predictions is less pronounced, highlighting a trade-off between attack effectiveness and stealthiness.

Concerning the Influence strategy, our NILS method exhibits a modest improvement over the LinkTeller baseline for the Twitch-FR dataset, as illustrated in Table \ref{tab:adv_strategies}. This suggests that the node injection property of our NILS attack is effective in this context. However, for the Twitch-RU dataset, NILS underperforms in comparison to the LinkTeller baseline. The most significant improvement is observed in the Flickr dataset, where the node injection property of NILS considerably increases the $F_1$ score from $0.32 \pm 0.13$ of LinkTeller to $0.89 \pm 0.10$. This outcome highlights the advantage of NILS attack's node injection method within the Influence strategy, particularly when compared to the LinkTeller attack, which employs the Influence strategy without node injection. 

These findings underscore the importance of considering both the effectiveness and stealthiness of malicious feature generation strategies when devising link inference attacks on GNNs.

\begin{table}[h]

\begin{adjustbox}{width=\columnwidth,center}
\centering

\begin{tabular}{lccc}
\toprule
Method & Twitch-FR & Twitch-RU & Flickr \\
\midrule
Class Rep. & $0.94 \pm 0.01$ & $0.83 \pm 0.06$ & $0.96 \pm 0.06$ \\
Max Attr.  & $0.99 \pm 0.00$  & $0.98 \pm 0.02$ & $\boldsymbol{1.00 \pm 0.00}$ \\
All-ones   & $\boldsymbol{0.99 \pm 0.00}$ & $\boldsymbol{0.97 \pm 0.01}$ & $0.99 \pm 0.02$ \\
All-zeros  & $0.58 \pm 0.02$  & $0.48 \pm 0.01$  & $0.71 \pm 0.07$ \\
Identity   & $0.81 \pm 0.02$  & $0.69 \pm 0.01$  & $0.95 \pm 0.07$ \\
Influence NILS  & $0.81 \pm 0.02$  & $0.70 \pm 0.01$  & $0.89 \pm 0.10$ \\
Influence LinkTeller \cite{linkteller}   & $0.80 \pm 0.02$  & $0.74 \pm 0.01$  & $0.32 \pm 0.13$ \\
\bottomrule
\end{tabular}
\end{adjustbox}
\caption{$F_1$ scores and standard deviations for different attack methods and datasets.}
\label{tab:adv_strategies}
\end{table}
\subsection{Comparison with the baselines}

In this study, we conducted experiments to evaluate the performance of our proposed NILS attack in comparison to the LinkTeller attack using the same experimental setup. Our focus is on analyzing the optimal attacks for both approaches, which involved accurately estimating the number of neighbors of the target set nodes. The results, summarized in Table \ref{tab:comp_LT}, demonstrate that our attack outperforms LinkTeller on both Twitch datasets (TWITCH-FR and TWITCH-RU). Furthermore, our method exhibits a substantial improvement over LinkTeller on the Flickr dataset, achieving nearly double the precision and recall values. Notably, our attack demonstrates stable performance across varying node degrees, with only a marginal decrease in effectiveness for high-degree target nodes. This can be attributed to the smaller influence that each neighboring node has on the aggregation of the GCN layer when the target node degree is high. Overall, our proposed NILS attack demonstrates consistently a superior performance compared to the LinkTeller attack.

We further compare our attack with link-stealing attacks introduced in \cite{he2021stealing}, where the authors' various attack strategies rely on different types of background knowledge available to the adversary, such as node attributes and shadow datasets. Specifically, in their Attack-2, the adversary has access to both the features and prediction scores of the nodes. Utilizing this information, the adversary creates two types of attacks: LSA2-attr and LSA2-post. LSA2-attr calculates distances between node attributes, while LSA2-post computes distances between node prediction scores (posteriors). It is important to highlight that these two attacks align closely with our threat model, as both assume that the adversary has access to the features and prediction scores of the target node. This similarity in assumptions renders these attacks particularly relevant for comparison with our proposed NILS attack. The attacks are executed under the transductive setting, where training and inference occur on the same graph. As shown in Table \ref{tab:LST_comp}, our proposed NILS attack outperforms the LSA2-post and LSA2-attr attacks constructed in \cite{he2021stealing}. However, our attack performance is nearly equivalent to that of LinkTeller. These results demonstrate that NILS attack maintains effectiveness under the transductive setting, just as in the inductive setting.
\begin{table*}[]
\begin{tabular}{cccccccc}
\toprule
\multirow{2}{*}{Dataset} &
  \multirow{2}{*}{Method} &
  \multicolumn{2}{c}{low} &
  \multicolumn{2}{c}{uncontrained} &
  \multicolumn{2}{c}{high} \\ \cline{3-8} 
 &
   &
  precision &
  recall &
  precision &
  recall &
  precision &
  recall \\ \hline
\multirow{2}{*}{TWITCH-FR} &
  NILS (Ours) &
  $100.0 \pm \scriptstyle 0.0$ &
  $100.0 \pm \scriptstyle 0.0$ &
  $99.13 \pm \scriptstyle 0.8$ &
  $99.57 \pm \scriptstyle 0.35$ &
  $99.91 \pm \scriptstyle 2.6$ &
  $100.0\pm \scriptstyle 0.0$ \\
 &
  LinkTeller &
  $92.5 \pm \scriptstyle 5.4$ &
  $92.5 \pm \scriptstyle 5.4$ &
  $84.1 \pm \scriptstyle 3.7$ &
  $78.2 \pm \scriptstyle 1.9$ &
  $83.2 \pm \scriptstyle 1.4$ &
  $80.6 \pm \scriptstyle 6.7$ \\ \hline
\multirow{2}{*}{TWITCH-RU} &
  NILS (Ours) &
  $100.0 \pm \scriptstyle 0.0$ &
  $100.0 \pm \scriptstyle 0.0$ &
  $96.45 \pm \scriptstyle 0.4 $ &
  $ 98.34\pm \scriptstyle 0.7$ &
  $99.77 \pm \scriptstyle 0.1$ &
  $ 99.37\pm \scriptstyle 0.1$ \\
 &
  LinkTeller &
  $78.8 \pm \scriptstyle 1.9$ &
  $ 92.6 \pm \scriptstyle 5.5 $ &
  $ 71.8\pm \scriptstyle 2.2$ &
  $78.5 \pm \scriptstyle 2.4$ &
  $ 89.7\pm \scriptstyle 1.7 $ &
  $65.7 \pm \scriptstyle 3.9 $ \\ \hline
\multirow{2}{*}{Flickr} &
  NILS (Ours) &
  $100.0\pm \scriptstyle 0.0$ &
  $100.0\pm \scriptstyle 0.0$ &
  $99.11\pm \scriptstyle 1.7$ &
  $95.83\pm \scriptstyle 5.0$ &
  $93.72\pm \scriptstyle 3.1$ &
  $78.9\pm \scriptstyle 1.9 $ \\
 &
  LinkTeller &
  $51.0 \pm \scriptstyle 7.0$ &
  $53.3\pm \scriptstyle 4.7$ &
  $33.8\pm \scriptstyle 13.3$ &
  $32.1\pm \scriptstyle 13.3$ &
  $18.2\pm \scriptstyle 4.5$ &
  $18.5\pm \scriptstyle 6.1$ \\ \hline
\end{tabular}
\caption{Comparative performance of our proposed attack NILS and LinkTeller across three datasets (TWITCH-FR, TWITCH-RU, and Flickr) under low, unconstrained, and high constraint settings. The results are presented in terms of precision and recall with corresponding standard deviations}
\label{tab:comp_LT}
\end{table*}

\begin{table}[]
\centering
\begin{adjustbox}{width=\columnwidth,center}
\begin{tabular}{ccccccc}
\toprule
\multirow{2}{*}{Method} &
  \multicolumn{2}{c}{Cora} &
  \multicolumn{2}{c}{Citeseer} &
  \multicolumn{2}{c}{Pubmed} \\ \cline{2-7} 
 &
  precision &
  recall &
  precision &
  recall &
  precision &
  recall \\ \midrule
NILS (Ours) &
  $ 99.7\pm \scriptstyle 0.2$ &
  $ 99.6\pm \scriptstyle 0.3 $ &
  $97.4 \pm \scriptstyle 0.2 $ &
  $98.2 \pm \scriptstyle 0.1$ &
  $ 99.7\pm \scriptstyle 0.0 $ &
  $100.0 \pm \scriptstyle 0.0 $ \\
LinkTeller &
  $99.5 \pm \scriptstyle 0.1 $ &
  $ 99.5\pm \scriptstyle 0.1$ &
  $99.7 \pm \scriptstyle 0.0$ &
  $99.7 \pm \scriptstyle 0.0$ &
  $99.7 \pm \scriptstyle 0.0$ &
  $99.7 \pm \scriptstyle 0.0$ \\
LSA2-post &
  $ 86.7 \pm \scriptstyle 0.2 $ &
  $ 86.7\pm \scriptstyle 0.2$ &
  $ 90.1 \pm \scriptstyle 0.2$ &
  $ 90.1 \pm \scriptstyle 0.2$ &
  $ 78.8\pm \scriptstyle 0.1$ &
  $ 78.8\pm \scriptstyle 0.1$ \\
LSA2-attr &
  $73.6 \pm \scriptstyle 0.1$ &
  $73.6 \pm \scriptstyle 0.1$ &
  $80.9 \pm \scriptstyle 0.1$ &
  $80.9 \pm \scriptstyle 0.1$ &
  $ 82.4\pm \scriptstyle0.1 $ &
  $ 82.4\pm \scriptstyle0.1 $ \\
\bottomrule
  
\end{tabular}
\end{adjustbox}
\caption{Comparative performance of NILS attack with LinkTeller \cite{linkteller} and link-stealing attacks in \cite{he2021stealing} across three datasets (Cora, Citeseer, and Pubmed).}
\label{tab:LST_comp}
\end{table}

\subsection{Depth of the GNN}
In this section, we examine the impact of increasing the depth of GNN on the success rate of the attack for the Twitch-Fr dataset. Our findings illustrated in Figure \ref{fig:depth_imact_LT} indicate that as the depth of the GNN increases, the attack's success rate decreases, which can be attributed to the dilution of the injected poisoning node's influence within the target node's neighborhood. As the GNN depth increases, the model aggregates information from a larger neighborhood, encompassing nodes that are $k-$hops away from the target node. Consequently, the injected malicious node's features become one among many contributing factors in the aggregated information, leading to a dilution of its influence. This reduction in the injected node's impact on the aggregated information diminishes the overall effectiveness of the attack, making it less successful in altering the predictions of the target node's neighbors.

In comparison with LinkTeller \cite{linkteller}, as shown in Table \ref{tab:depth_imact_LT}, NILS outperforms LinkTeller \cite{linkteller} across various GCN depths. Specifically, for Twitch-FR dataset, NILS demonstrates higher precision and recall values when the GCN depth is 3 (precision: $85.06 \pm \scriptstyle 1.2$, recall: $81.56 \pm \scriptstyle 1.2$) compared to the LinkTeller method (precision: $50.01 \pm \scriptstyle 5.1$, recall: $46.6 \pm \scriptstyle 5.0$). Notably, NILS consistently outperforms LinkTeller even when comparing the attack performance of LinkTeller with a GCN depth of 2 and NILS with a GCN depth of 3. Specifically, for Twitch-FR dataset, NILS demonstrates higher precision and recall values at a GCN depth of 3 (precision: $85.06 \pm \scriptstyle 1.2$, recall: $81.56 \pm \scriptstyle 1.2$) compared to the LinkTeller method with a GCN depth of 2 (precision: $84.1 \pm \scriptstyle 3.7$, recall: $78.2 \pm \scriptstyle 1.9$). These results highlight the effectiveness of our node injection strategy, as it consistently outperforms the LinkTeller method across different depths of the GCN.

\begin{figure}[ht]
  \centering
  \includegraphics[width=\columnwidth]{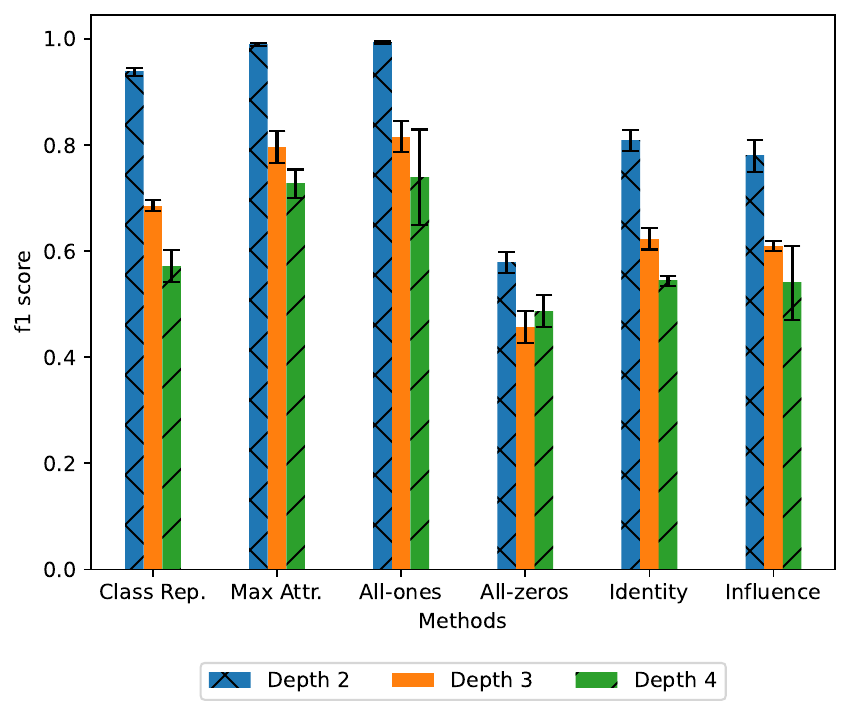}
  \caption{Success rates of the attack for different depths and malicious features generation strategies for Twitch-FR dataset}
  \label{fig:depth_imact_LT}
\end{figure}

\begin{table}[]
\begin{adjustbox}{width=\columnwidth,center}
\begin{tabular}{cccccc}
\toprule
\multirow{2}{*}{Dataset} &
  \multirow{2}{*}{Method} &
  \multicolumn{2}{c}{Depth-2} &
  \multicolumn{2}{c}{Depth-3} \\ \cline{3-6} 
 &
   &
  precision &
  recall &
  precision &
  recall \\ \midrule
\multirow{2}{*}{TWITCH-FR} & NILS (Ours) & $99.13 \pm \scriptstyle 0.8$ & $99.57 \pm \scriptstyle 0.35$ & $85.06\pm \scriptstyle 1.2$   & $81.56 \pm \scriptstyle 1.2$ \\
 &
  LinkTeller &
  $84.1 \pm \scriptstyle 3.7$ &
  $78.2 \pm \scriptstyle 1.9$ &
  $50.1 \pm \scriptstyle 5.1$ &
  $46.6 \pm \scriptstyle 5.0$ \\ \midrule
\multirow{2}{*}{TWITCH-RU} & NILS (Ours) & $96.45 \pm \scriptstyle 0.4$ & $98.34 \pm \scriptstyle 0.7$  & $78.78 \pm \scriptstyle 3.8 $ & $ 76.35\pm \scriptstyle 9.3$ \\
 &
  LinkTeller &
  $71.8\pm \scriptstyle 2.2$ &
  $78.5 \pm \scriptstyle 2.4 $ &
  $45.7\pm \scriptstyle 2.2$ &
  $50.0 \pm \scriptstyle 2.8$ \\
\bottomrule
\end{tabular}
\end{adjustbox}
  \caption{Success rates of the attack for different depths in comparison with LinkTeller \cite{linkteller}. We use the all-ones strategy and Twitch-FR dataset.}
  \label{tab:depth_imact_LT}
\end{table}

\section{Defense\label{sec:defense}}

This section introduces the basic notions of DP in the context of GNNs. As a reminder, the goal is to protect the privacy of the graph, in the sense of preventing an adversary from discovering whether, in a given graph, there is a link between two nodes. With this aim, we need to define the neighbouring relation of graphs and further revise the definition of DP.

\subsection{DP for graphs}
Recall from Sec.~\ref{sec:background:DP} that the notion of neighborhood of DP was defined originally for microdata, and, 
accordingly, two databases are said to be neighbors if they differ just in one record.
In the context of graphs at hand, however, this notion must be adapted since two graphs may differ with respect to either one edge or one node.

In the literature, we find two attempts~\cite{kossinets2006empirical,hay2009accurate} to adapt DP to graphs. 
Before examining them, recall from Sec.~\ref{sec:GNNsOverview} that a graph $\mathcal{G} =(V,E)$ is represented with an adjacency matrix $A$, whereby $A_{ij}=1$ if there is a link between node $i$ and node $j$, and $A_{ij} = 0$ otherwise (where $i,j \in \{1,\ldots,|V|\}$). 

\begin{definition}[\textbf{Edge-level adjacent graphs} \cite{kossinets2006empirical}]
$\mathcal{G}$ and $\mathcal{G'}$ are considered \emph{edge-level adjacent graphs} if one can be obtained from the other by removing a single edge. In other words, $\mathcal{G}$ and $\mathcal{G'}$ differ by at most one edge. Hence, their adjacency matrices differ by one element only.
\end{definition}
Accordingly, an edge-level DP mechanism is defined as follows:
\begin{definition}[\textbf{$(\varepsilon, \delta)$-Edge-level differential privacy}]
	\label{def:edgedp}
	A randomized mechanism $\mathcal{M}$ satisfies \textbf{$(\varepsilon,\delta)$-edge-level DP} with $\varepsilon,\delta \geqslant 0$ if, for all pairs of
	edge-level adjacent graphs $\mathcal{G},\mathcal{G}'$ and for all measurable $\mathcal{O}\subseteq \Range(\M)$,
	
	$$\oP\{\mathcal{M}(\mathcal{G})\in \mathcal{O}\} \leqslant e^{\varepsilon} \oP\{\mathcal{M}(\mathcal{G}')\in \mathcal{O}\} + \delta.$$
\end{definition}

\begin{definition}[\textbf{Node-level adjacent graphs} \cite{hay2009accurate}]
\label{def:node-level}
$\mathcal{G}$ and $\mathcal{G'}$ are said to be \emph{node-level adjacent graphs} if one can be obtained from the other by removing a single node and all of its incident edges.
\end{definition}
Node-level DP is defined analogously as follows:
\begin{definition}[\textbf{$(\varepsilon, \delta)$-Node-level differential privacy}]
	\label{def:nodedp}
	A randomized mechanism $\mathcal{M}$ satisfies \textbf{$(\varepsilon,\delta)$-node-level DP} with $\varepsilon,\delta \geqslant 0$ if, for all pairs of node-level adjacent graphs $\mathcal{G},\mathcal{G}'$ and for all measurable $\mathcal{O}\subseteq\Range(\mathcal{M})$, the following inequality holds:
	
	$$\oP\{\mathcal{M}(\mathcal{G})\in \mathcal{O}\} \leqslant e^{\varepsilon} \oP\{\mathcal{M}(\mathcal{G}')\in \mathcal{O}\} + \delta$$
\end{definition}

\subsection{One-node-one-edge-level DP}
The adversary defined in Sec.~\ref{sec:node-injection-attack} adds a malicious node to a graph and connects it to a target node through a \emph{single} edge. Countering such an adversary with a node-level DP mechanism (see Definition~\ref{def:node-level}) is clearly not a suitable choice in terms of model accuracy since node-level DP targets a stronger adversary. Trying to hide the presence or absence of one node and \emph{all} of its incident edges intuitively would increase the scale of the noise to be added (for example to the original adjacency matrix), and incur more data inaccuracy than necessary.
Motivated by this, we define a new notion of neighboring graphs (and the corresponding DP mechanism), which is designed to specifically counter the adversary proposed in this work. 
\begin{definition}[\textbf{One-node-one-edge-level adjacent graphs}]
\label{def:1n1e-graphs}
$\mathcal{G}$ and $\mathcal{G'}$ are considered \emph{one-node-one-edge-level adjacent graphs} if one can be obtained from the other by adding a single node with one edge only.
\end{definition}
Note that, as in node-level adjacent graphs (Definition~\ref{def:node-level}), the adjacency matrices of two neighboring graphs (in the sense of one-node-one-edge) differs by one row and one column only, but unlike node-level, the difference in $L_1-$norm between the adjacency matrices is always one.
Based on Definition~\ref{def:1n1e-graphs}, a one-node-one-edge-level DP is defined as follows:
\begin{definition}[\textbf{$(\varepsilon, \delta)$-One-node-one-edge-level differential privacy}]
	\label{def:nodedp}
	A randomized mechanism $\mathcal{M}$ satisfies \textbf{$(\varepsilon,\delta)$-one-node-one-edge-level DP} with $\varepsilon,\delta \geqslant 0$ if, for all pairs of one-node-one-edge-level adjacent graphs $\mathcal{G},\mathcal{G}'$ and for all measurable $\mathcal{O}\subseteq \Range(\mathcal{M})$,
the following holds:
	$$\oP\{\mathcal{M}(\mathcal{G})\in \mathcal{O}\} \leqslant e^{\varepsilon} \oP\{\mathcal{M}(\mathcal{G}')\in \mathcal{O}\} + \delta$$
\end{definition}

\subsection{Countermeasures for our attack}

In this section, we describe one DP-based strategy, namely the LapGraph mechanism, which was introduced in \cite{linkteller}. A much simpler defense approach against any privacy attack to GNNs would of course be output perturbation~\cite{outputperturbation}, whereby the very same output of the GNN prediction is perturbed with some DP mechanism (e.g., the classical Laplace mechanism). While this solution is straightforward to implement and indeed can be used to satisfy the one-node-one-edge-level DP notion, unfortunately, it would significantly deteriorate the accuracy of the GNN output. 
It is easy to see that the $L_1$-global sensitivity of a prediction matrix for the set of nodes $V_{\mathcal{A}}$ is as large as $2\,|V_{\mathcal{A}}|$, which makes us rule out output perturbation. 

To defend against the newly proposed attack, similar to \cite{linkteller}, we propose to apply the LapGraph algorithm, which consists in perturbing the adjacency matrix using the Laplace mechanism and binarizing it by replacing the top-$N$ largest values by 1 and the remaining values by 0. Here, $N$ represents the estimated number of edges in the graph, which is also computed using the Laplace mechanism. 

By leveraging the post-processing property of DP\footnote{The post-processing of DP allows arbitrary data-independent transformations to DP outputs without affecting their privacy guarantee~\cite{postprocessing}.}, the edge information remains protected even if the adversary observes the predictions generated by the GNN. Furthermore, 
 
each time a user connects a new node, a new adjacency matrix is generated following the same LapGraph mechanism, accumulating this way the privacy budget by the sequential composition property of DP \cite{dwork2014algorithmic}.

Although the LapGraph mechanism was proposed to meet edge-level DP, it is not difficult to show that the mechanism can also be used to satisfy one-node-one-edge-level DP. 
For this, let $f_A$ be the query function returning the adjacency matrix of a graph $G$. Unlike edge-level neighborhood, the corresponding matrices $A, A'$ of two one-node-one-edge neighboring graphs $G,G'$ have different dimensions, namely, 
either $A\in \mathbb{R}^{n \times n}$ and $A'\in \mathbb{R}^{(n+1) \times (n+1)}$, or $A\in \mathbb{R}^{(n+1) \times (n+1)}$ and $A'\in \mathbb{R}^{n \times n}$. Without loss of generality, we assume the former case, where $A$ and $A'$ represent the adjacency matrices \emph{before} and \emph{after} the new node is connected to $G$ (resulting in $G'$). We shall also assume that the new node corresponds to the $(n+1)$-th row and, for symmetry, to the $(n+1)$-th column of $A'$. Precisely, since any adjacency matrix is symmetric by definition, the computation of the sensitivity of $f_A$ only requires the upper or lower triangular matrix of $A$.

To enable the subtraction operation $A-A'$ implicit in the definition of the global sensitivity (see Definition~\ref{def:GS}), we append one zero-row and one-zero column to $A$ and denote the resulting matrix by $\bar{A}\in \mathbb{R}^{(n+1) \times (n+1)}$.
As in the case of $A'$, we assume that the appended row and column are in the 
$(n+1)$-th position of $\bar{A}$.

To compute the sensitivity of $f_A$ for the notion of one-node-one-edge neighboring graphs, we just need to note that the $(n+1)$-th columns (or rows, if we consider the lower triangular of the adjacency matrix) of $\bar{A}$ and $A'$ always differ in one element. The reason is because one-node-one-edge neighboring graphs differ in only one edge. As a result,
$$\|\bar{A}-A'\|_1=1$$
for any pair of neighboring graphs,
which yields an $L_1$-global sensitivity of 1, as in the original LapGraph mechanism intended for edge-level DP. The fact that the two sensitivities coincide implies that the LapGraph version utilized in this work will provide stronger protection for the same level of utility, compared to the original LapGraph.
The reason for that is because, while the scale of the Laplace noise will be the same for a same $\varepsilon$, one-node-one-edge guarantees indistinguishability between any pair of graphs differing not only in one edge and but also in one node. 
\begin{figure*}[t]
  \centering
  \begin{subfigure}[b]{0.3\textwidth}
    \includegraphics[width=\textwidth]{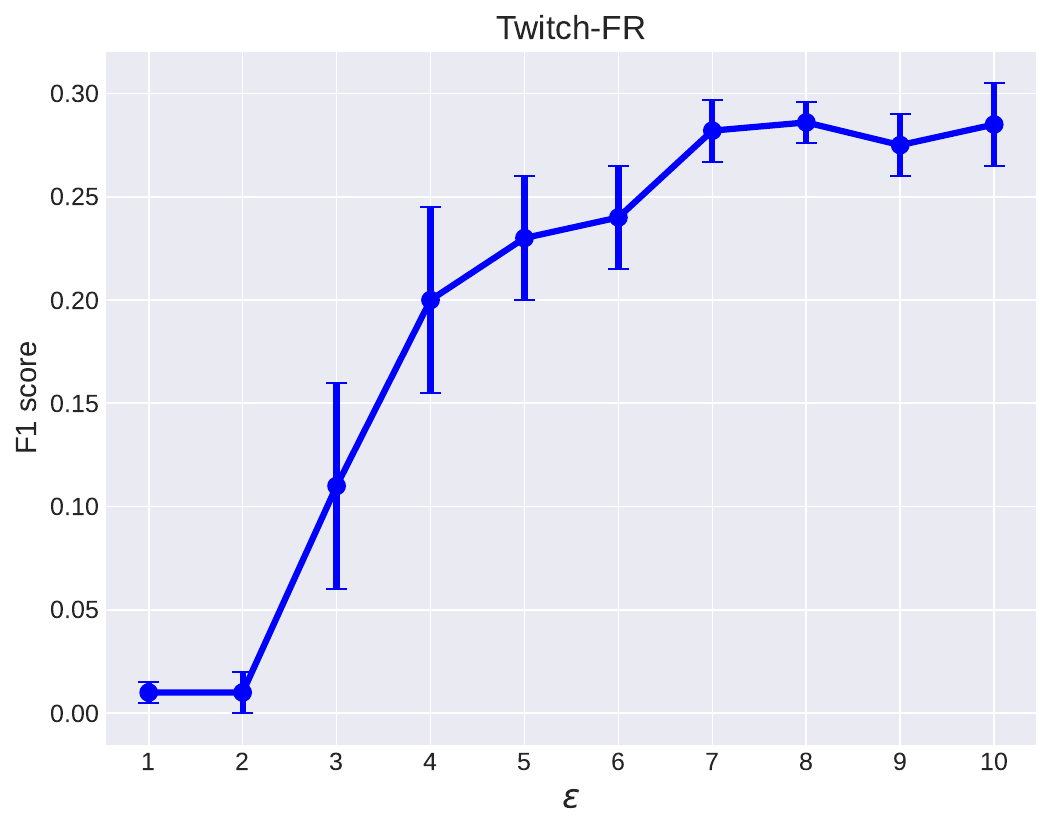}
  \end{subfigure}
  \hfill
  \begin{subfigure}[b]{0.3\textwidth}
    \includegraphics[width=\textwidth]{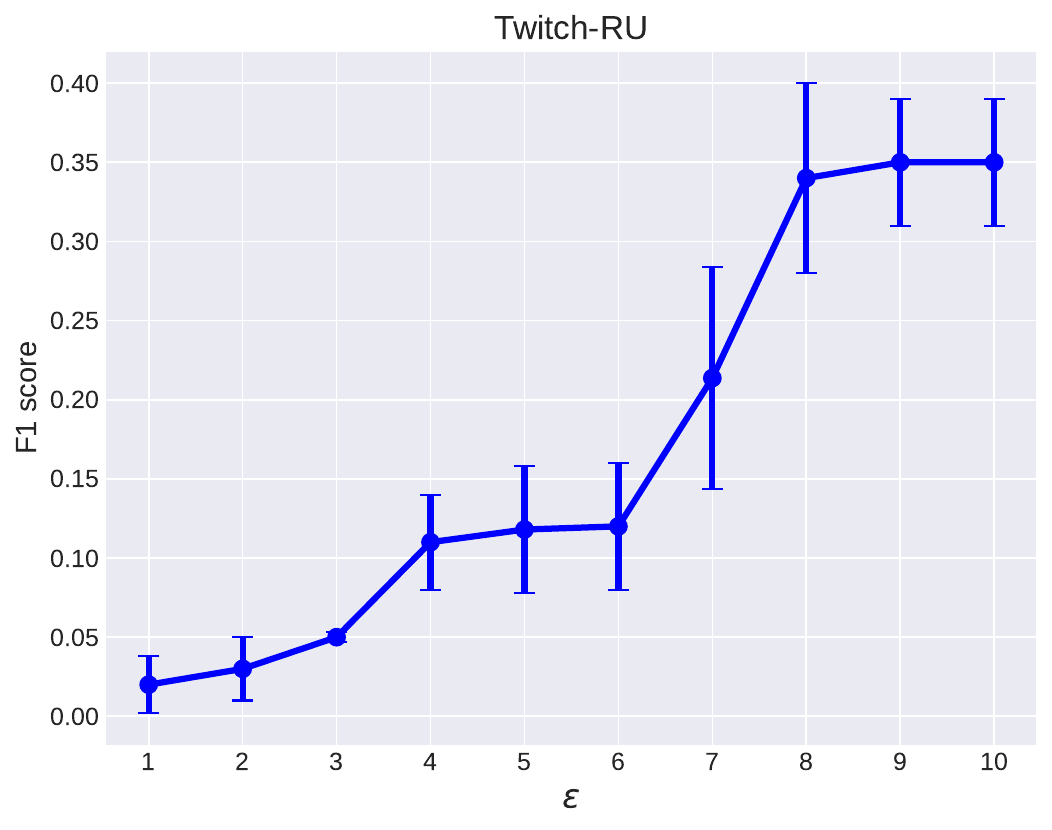}
  \end{subfigure}
  \hfill
  \begin{subfigure}[b]{0.3\textwidth}
    \includegraphics[width=\textwidth]{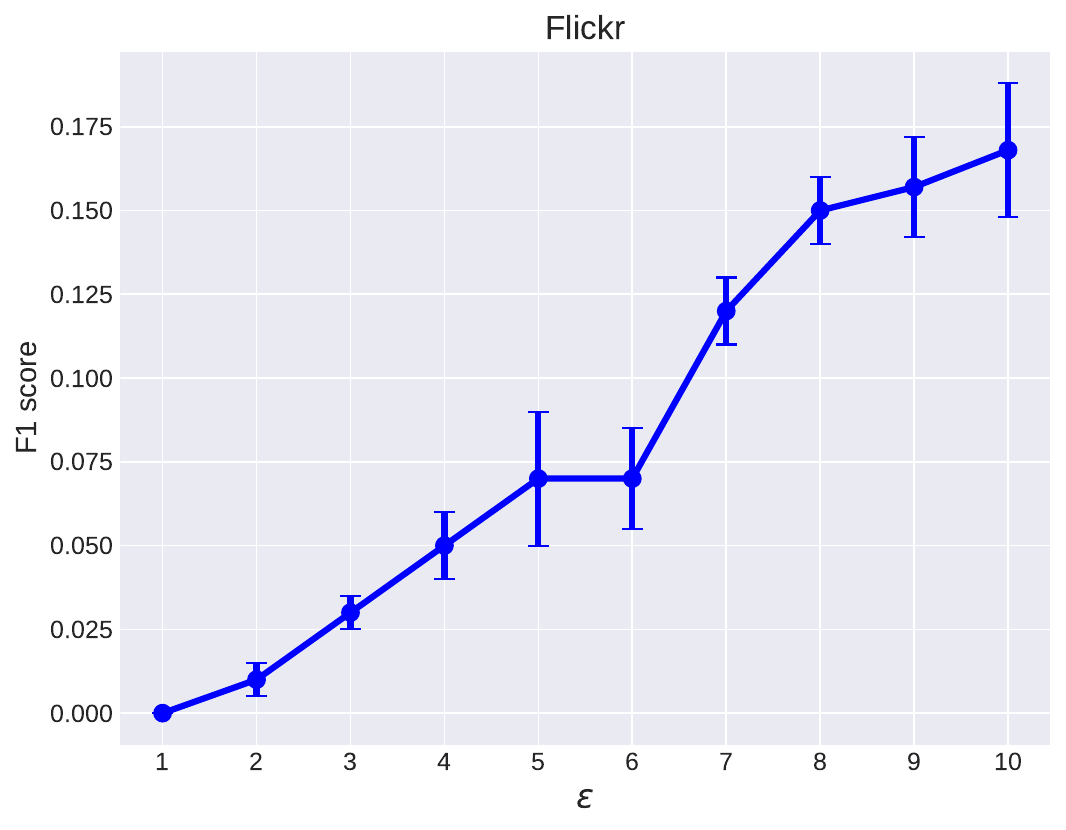}
  \end{subfigure}
  \caption{$F_1$ score of the attack for different values of $\varepsilon$.}
  \label{fig:lapgraph_effectiveness}
\end{figure*}

\subsection{LapGraph evaluation}
In this section, we evaluate the effectiveness of LapGraph \cite{linkteller} in reducing the success of  NILS attack while ensuring our one-node-one-edge-level DP notion. We also investigate the utility of GCN models trained with LapGraph protection.

\subsubsection{Evaluation setup:} We use the same training hyperparameters and normalization techniques as in the vanilla case, where DP is not applied. Initially, we protect the training graph with LapGraph. Following that, we apply LapGraph each time the graph changes due to node injection by the adversary.
In line with the setup in \cite{linkteller}, we compute the $F_1$ score for our NILS attack as well as the classification task's $F_1$ score for the GCN. This allows us to measure LapGraph protection along with the GCN utility across various privacy budgets $\varepsilon$. We report the results averaged over 5 runs with different random seeds for LapGraph.

\subsubsection{Evaluation results:} Figure \ref{fig:lapgraph_effectiveness} presents the $F_1$ score of the attack for various $\varepsilon$ values. We observe that applying LapGraph reduces the effectiveness of NILS. The $F_1$ score becomes almost zero when the privacy budget $\varepsilon$ is small. However, for large $\varepsilon$, LapGraph provides moderate protection, but the attack's $F_1$ score remains significantly lower than in the non-private case where DP is not applied.

For comparison, in the LinkTeller \cite{linkteller} attack, where LapGraph is applied only once to ensure edge-level DP, LapGraph offers limited protection when $\varepsilon$ is large, allowing LinkTeller to achieve a success rate nearly as high in the non-private case. Conversely, in our scenario, where LapGraph is also applied after the adversary's node injection, LapGraph provides stronger protection. The application of LapGraph during inference makes it more challenging for the adversary to distinguish between the target node's neighbors and non-neighbors, as the prediction scores of all target nodes change after each inference query.
Consequently, the distances between the prediction scores $P$ and $P'$, before and after the node injection, become noisier due to LapGraph's application following the node injection.

To provide insights about the privacy-utility tradeoff of LapGraph, we present in Figure \ref{fig:lapgraph_utility} the utility of the GCNs for different values of the privacy budget. We observe that the utility increases when $\varepsilon$ increases, as expected. Large values of $\varepsilon \geq 7$ give a better utility close to that in the non-private vanilla case. Therefore, carefully choosing an $\varepsilon$ will give fairly good utility and a certain level of protection against NILS attack.

\begin{figure*}[t]
  \centering
  \begin{subfigure}[b]{0.3\textwidth}
    \includegraphics[width=\textwidth]{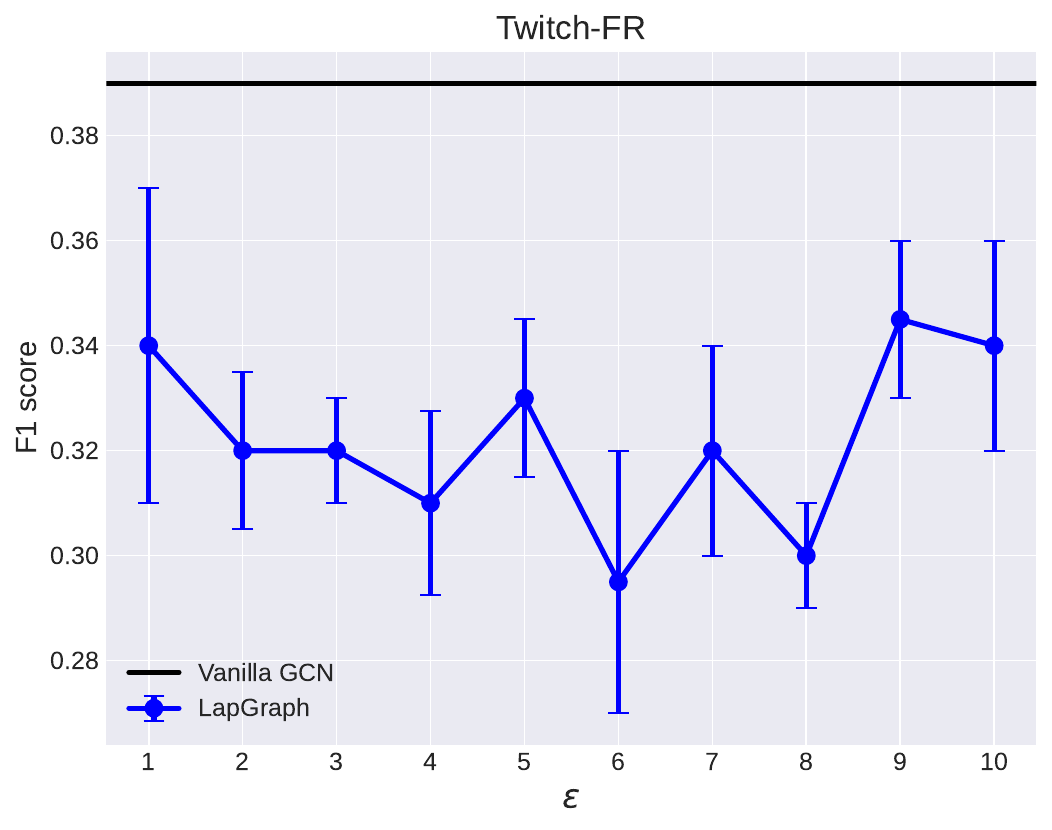}
  \end{subfigure}
  \hfill
  \begin{subfigure}[b]{0.3\textwidth}
    \includegraphics[width=\textwidth]{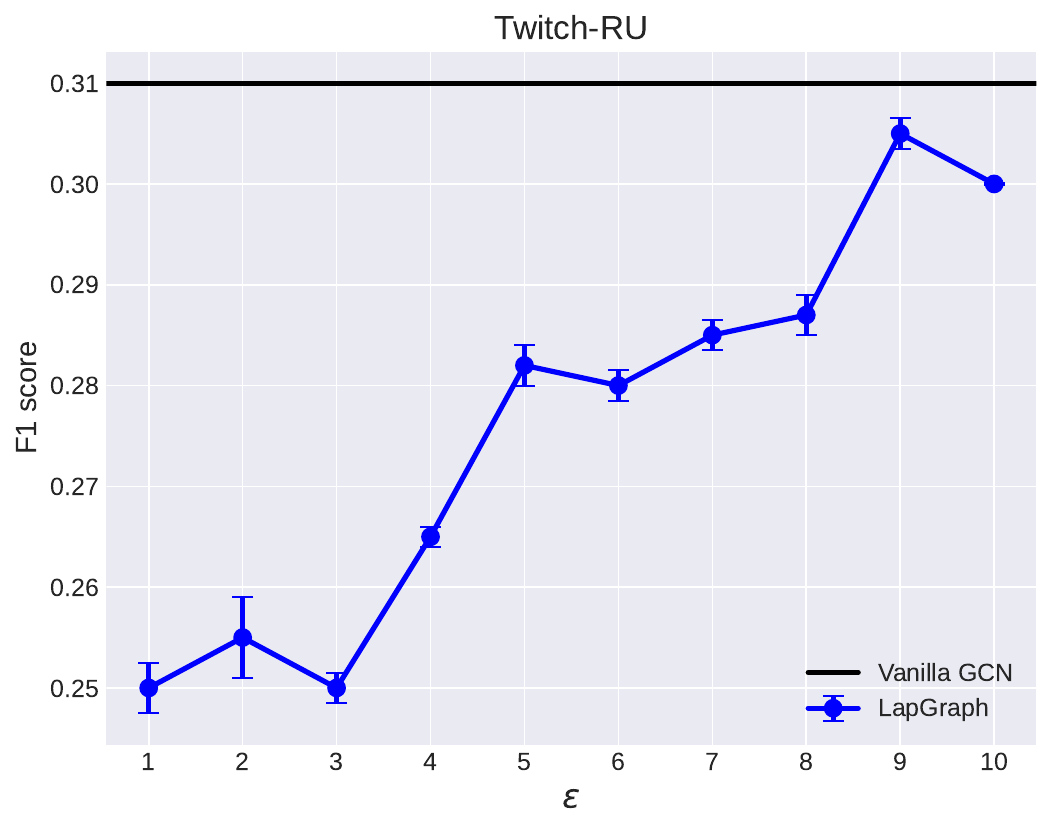}
  \end{subfigure}
  \hfill
  \begin{subfigure}[b]{0.3\textwidth}
    \includegraphics[width=\textwidth]{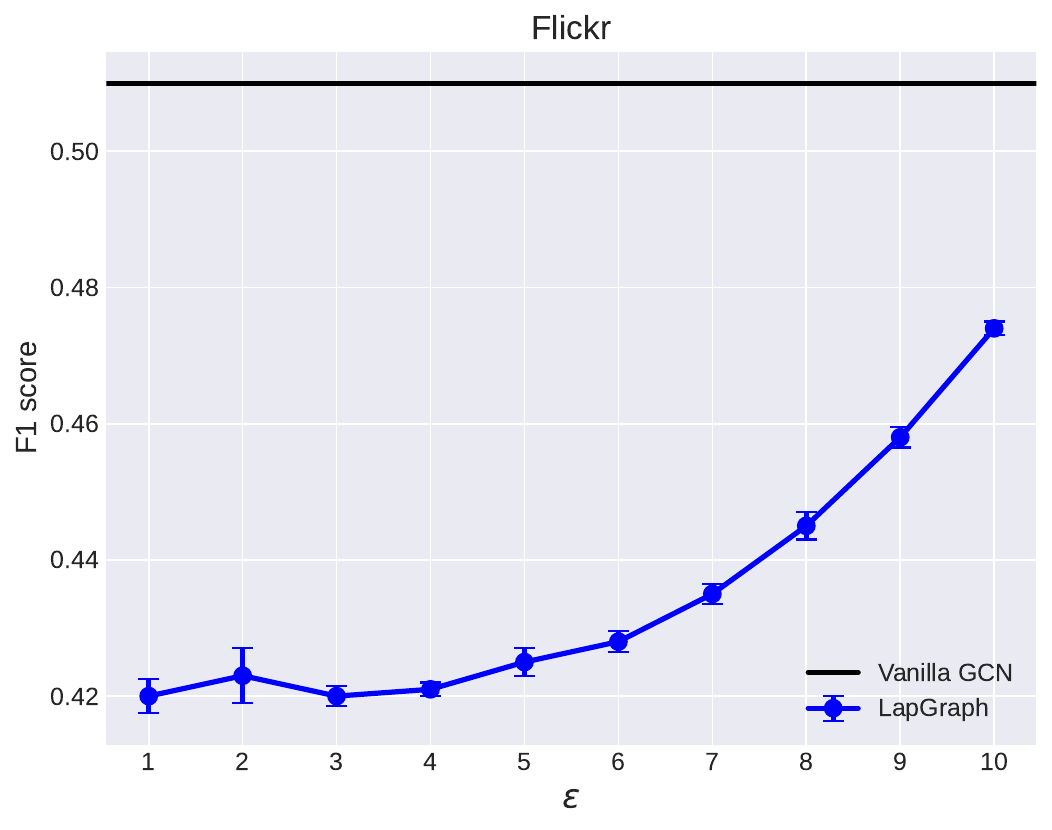}
  \end{subfigure}
  \caption{$F_1$ score utility of the GCN for different values of $\varepsilon$.}
  \label{fig:lapgraph_utility}
\end{figure*}

\section{Conclusion\label{sec:conclusion}}
In this paper, we have presented a powerful new NILS attack—a link-stealing attack using node injection against GNNs. Our results have demonstrated the superior performance of NILS compared to previous attacks, further emphasizing the vulnerabilities of GNNs regarding edge information leakage. We have also evaluated NILS against differentially private GNNs, ensuring a one-node-one-edge-level DP notion specifically designed to protect against our proposed attack.

\newpage
\bibliographystyle{ACM-Reference-Format}
\bibliography{sample-base}


\begin{thebibliography}{41}


\ifx \showCODEN    \undefined \def \showCODEN     #1{\unskip}     \fi
\ifx \showDOI      \undefined \def \showDOI       #1{#1}\fi
\ifx \showISBNx    \undefined \def \showISBNx     #1{\unskip}     \fi
\ifx \showISBNxiii \undefined \def \showISBNxiii  #1{\unskip}     \fi
\ifx \showISSN     \undefined \def \showISSN      #1{\unskip}     \fi
\ifx \showLCCN     \undefined \def \showLCCN      #1{\unskip}     \fi
\ifx \shownote     \undefined \def \shownote      #1{#1}          \fi
\ifx \showarticletitle \undefined \def \showarticletitle #1{#1}   \fi
\ifx \showURL      \undefined \def \showURL       {\relax}        \fi
\providecommand\bibfield[2]{#2}
\providecommand\bibinfo[2]{#2}
\providecommand\natexlab[1]{#1}
\providecommand\showeprint[2][]{arXiv:#2}

\bibitem[Atwood and Towsley(2016)]%
        {gnn1}
\bibfield{author}{\bibinfo{person}{James Atwood} {and} \bibinfo{person}{Don
  Towsley}.} \bibinfo{year}{2016}\natexlab{}.
\newblock \showarticletitle{Diffusion-Convolutional Neural Networks}. In
  \bibinfo{booktitle}{\emph{Proceedings of the 30th International Conference on
  Neural Information Processing Systems}} (Barcelona, Spain)
  \emph{(\bibinfo{series}{NIPS'16})}. \bibinfo{publisher}{Curran Associates
  Inc.}, \bibinfo{address}{Red Hook, NY, USA}, \bibinfo{pages}{2001–2009}.
\newblock
\showISBNx{9781510838819}


\bibitem[Brunet et~al\mbox{.}(2016)]%
        {node_dp_Brunet}
\bibfield{author}{\bibinfo{person}{Solenn Brunet},
  \bibinfo{person}{S{\'{e}}bastien Canard}, \bibinfo{person}{S{\'{e}}bastien
  Gambs}, {and} \bibinfo{person}{Baptiste Olivier}.}
  \bibinfo{year}{2016}\natexlab{}.
\newblock \showarticletitle{Novel differentially private mechanisms for
  graphs}.
\newblock \bibinfo{journal}{\emph{{IACR} Cryptol. ePrint Arch.}}
  (\bibinfo{year}{2016}), \bibinfo{pages}{745}.
\newblock
\urldef\tempurl%
\url{http://eprint.iacr.org/2016/745}
\showURL{%
\tempurl}


\bibitem[Cadwalladr and Graham-Harrison(2018)]%
        {camb_analytica}
\bibfield{author}{\bibinfo{person}{Carole Cadwalladr} {and}
  \bibinfo{person}{Emma Graham-Harrison}.} \bibinfo{year}{2018}\natexlab{}.
\newblock \showarticletitle{Revealed: 50 million Facebook profiles harvested
  for Cambridge Analytica in major data breach}.
\newblock \bibinfo{journal}{\emph{The Guardian}} (\bibinfo{date}{17 Mar}
  \bibinfo{year}{2018}).
\newblock
\urldef\tempurl%
\url{https://www.theguardian.com/news/2018/mar/17/cambridge-analytica-facebook-influence-us-election}
\showURL{%
\tempurl}


\bibitem[Chen et~al\mbox{.}(2014)]%
        {Rui_adj_release}
\bibfield{author}{\bibinfo{person}{Rui Chen}, \bibinfo{person}{Benjamin C.~M.
  Fung}, \bibinfo{person}{Philip~S. Yu}, {and} \bibinfo{person}{Bipin~C.
  Desai}.} \bibinfo{year}{2014}\natexlab{}.
\newblock \showarticletitle{Correlated network data publication via
  differential privacy}.
\newblock \bibinfo{journal}{\emph{{VLDB} J.}} \bibinfo{volume}{23},
  \bibinfo{number}{4} (\bibinfo{year}{2014}), \bibinfo{pages}{653--676}.
\newblock
\urldef\tempurl%
\url{https://doi.org/10.1007/s00778-013-0344-8}
\showDOI{\tempurl}


\bibitem[Conti et~al\mbox{.}(2022)]%
        {conti2022label_MIA}
\bibfield{author}{\bibinfo{person}{Mauro Conti}, \bibinfo{person}{Jiaxin Li},
  \bibinfo{person}{Stjepan Picek}, {and} \bibinfo{person}{Jing Xu}.}
  \bibinfo{year}{2022}\natexlab{}.
\newblock \showarticletitle{Label-Only Membership Inference Attack against
  Node-Level Graph Neural Networks}. In \bibinfo{booktitle}{\emph{Proceedings
  of the 15th ACM Workshop on Artificial Intelligence and Security}}.
  \bibinfo{pages}{1--12}.
\newblock


\bibitem[Dai et~al\mbox{.}(2022)]%
        {dai2022comprehensive_survey}
\bibfield{author}{\bibinfo{person}{Enyan Dai}, \bibinfo{person}{Tianxiang
  Zhao}, \bibinfo{person}{Huaisheng Zhu}, \bibinfo{person}{Junjie Xu},
  \bibinfo{person}{Zhimeng Guo}, \bibinfo{person}{Hui Liu},
  \bibinfo{person}{Jiliang Tang}, {and} \bibinfo{person}{Suhang Wang}.}
  \bibinfo{year}{2022}\natexlab{}.
\newblock \showarticletitle{A comprehensive survey on trustworthy graph neural
  networks: Privacy, robustness, fairness, and explainability}.
\newblock \bibinfo{journal}{\emph{arXiv preprint arXiv:2204.08570}}
  (\bibinfo{year}{2022}).
\newblock


\bibitem[Daigavane et~al\mbox{.}(2021)]%
        {node_dp_Daigavane}
\bibfield{author}{\bibinfo{person}{Ameya Daigavane}, \bibinfo{person}{Gagan
  Madan}, \bibinfo{person}{Aditya Sinha}, \bibinfo{person}{Abhradeep~Guha
  Thakurta}, \bibinfo{person}{Gaurav Aggarwal}, {and} \bibinfo{person}{Prateek
  Jain}.} \bibinfo{year}{2021}\natexlab{}.
\newblock \showarticletitle{Node-Level Differentially Private Graph Neural
  Networks}.
\newblock \bibinfo{journal}{\emph{CoRR}}  \bibinfo{volume}{abs/2111.15521}
  (\bibinfo{year}{2021}).
\newblock
\showeprint[arXiv]{2111.15521}
\urldef\tempurl%
\url{https://arxiv.org/abs/2111.15521}
\showURL{%
\tempurl}


\bibitem[Day et~al\mbox{.}(2016)]%
        {WeiYen_degree_dist}
\bibfield{author}{\bibinfo{person}{Wei{-}Yen Day}, \bibinfo{person}{Ninghui
  Li}, {and} \bibinfo{person}{Min Lyu}.} \bibinfo{year}{2016}\natexlab{}.
\newblock \showarticletitle{Publishing Graph Degree Distribution with Node
  Differential Privacy}. In \bibinfo{booktitle}{\emph{Proceedings of the 2016
  International Conference on Management of Data, {SIGMOD} Conference 2016, San
  Francisco, CA, USA, June 26 - July 01, 2016}},
  \bibfield{editor}{\bibinfo{person}{Fatma {\"{O}}zcan},
  \bibinfo{person}{Georgia Koutrika}, {and} \bibinfo{person}{Sam Madden}}
  (Eds.). \bibinfo{publisher}{{ACM}}, \bibinfo{pages}{123--138}.
\newblock
\urldef\tempurl%
\url{https://doi.org/10.1145/2882903.2926745}
\showDOI{\tempurl}


\bibitem[Defferrard et~al\mbox{.}(2016)]%
        {gnn3}
\bibfield{author}{\bibinfo{person}{Micha\"{e}l Defferrard},
  \bibinfo{person}{Xavier Bresson}, {and} \bibinfo{person}{Pierre
  Vandergheynst}.} \bibinfo{year}{2016}\natexlab{}.
\newblock \showarticletitle{Convolutional Neural Networks on Graphs with Fast
  Localized Spectral Filtering}. In \bibinfo{booktitle}{\emph{Advances in
  Neural Information Processing Systems}},
  \bibfield{editor}{\bibinfo{person}{D.~Lee}, \bibinfo{person}{M.~Sugiyama},
  \bibinfo{person}{U.~Luxburg}, \bibinfo{person}{I.~Guyon}, {and}
  \bibinfo{person}{R.~Garnett}} (Eds.), Vol.~\bibinfo{volume}{29}.
  \bibinfo{publisher}{Curran Associates, Inc.}
\newblock
\urldef\tempurl%
\url{https://proceedings.neurips.cc/paper_files/paper/2016/file/04df4d434d481c5bb723be1b6df1ee65-Paper.pdf}
\showURL{%
\tempurl}


\bibitem[Duddu et~al\mbox{.}(2021)]%
        {linkstealing_reconstruction}
\bibfield{author}{\bibinfo{person}{Vasisht Duddu}, \bibinfo{person}{Antoine
  Boutet}, {and} \bibinfo{person}{Virat Shejwalkar}.}
  \bibinfo{year}{2021}\natexlab{}.
\newblock \showarticletitle{Quantifying Privacy Leakage in Graph Embedding}. In
  \bibinfo{booktitle}{\emph{MobiQuitous 2020 - 17th EAI International
  Conference on Mobile and Ubiquitous Systems: Computing, Networking and
  Services}} (Darmstadt, Germany) \emph{(\bibinfo{series}{MobiQuitous '20})}.
  \bibinfo{publisher}{Association for Computing Machinery},
  \bibinfo{address}{New York, NY, USA}, \bibinfo{pages}{76–85}.
\newblock
\showISBNx{9781450388405}
\urldef\tempurl%
\url{https://doi.org/10.1145/3448891.3448939}
\showDOI{\tempurl}


\bibitem[Dwork(2006)]%
        {Dwork06A}
\bibfield{author}{\bibinfo{person}{Cynthia Dwork}.}
  \bibinfo{year}{2006}\natexlab{}.
\newblock \showarticletitle{Differential Privacy}. In
  \bibinfo{booktitle}{\emph{Automata, Languages and Programming, 33rd
  International Colloquium, {ICALP} 2006, Venice, Italy, July 10-14, 2006,
  Proceedings, Part {II}}} \emph{(\bibinfo{series}{Lecture Notes in Computer
  Science}, Vol.~\bibinfo{volume}{4052})},
  \bibfield{editor}{\bibinfo{person}{Michele Bugliesi}, \bibinfo{person}{Bart
  Preneel}, \bibinfo{person}{Vladimiro Sassone}, {and} \bibinfo{person}{Ingo
  Wegener}} (Eds.). \bibinfo{publisher}{Springer}, \bibinfo{pages}{1--12}.
\newblock
\urldef\tempurl%
\url{https://doi.org/10.1007/11787006\_1}
\showDOI{\tempurl}


\bibitem[Dwork et~al\mbox{.}(2006)]%
        {outputperturbation}
\bibfield{author}{\bibinfo{person}{Cynthia Dwork}, \bibinfo{person}{Frank
  McSherry}, \bibinfo{person}{Kobbi Nissim}, {and} \bibinfo{person}{Adam
  Smith}.} \bibinfo{year}{2006}\natexlab{}.
\newblock \showarticletitle{Calibrating Noise to Sensitivity in Private Data
  Analysis}. In \bibinfo{booktitle}{\emph{Theory of Cryptography}},
  \bibfield{editor}{\bibinfo{person}{Shai Halevi} {and} \bibinfo{person}{Tal
  Rabin}} (Eds.). \bibinfo{publisher}{Springer Berlin Heidelberg},
  \bibinfo{address}{Berlin, Heidelberg}, \bibinfo{pages}{265--284}.
\newblock


\bibitem[Dwork and Roth(2014)]%
        {dwork2014algorithmic}
\bibfield{author}{\bibinfo{person}{Cynthia Dwork} {and} \bibinfo{person}{Aaron
  Roth}.} \bibinfo{year}{2014}\natexlab{}.
\newblock \showarticletitle{The Algorithmic Foundations of Differential
  Privacy}.
\newblock \bibinfo{journal}{\emph{Found. Trends Theor. Comput. Sci.}}
  \bibinfo{volume}{9}, \bibinfo{number}{3-4} (\bibinfo{year}{2014}),
  \bibinfo{pages}{211--407}.
\newblock
\urldef\tempurl%
\url{https://doi.org/10.1561/0400000042}
\showDOI{\tempurl}


\bibitem[Hamilton et~al\mbox{.}(2017)]%
        {Hamilton_node_class}
\bibfield{author}{\bibinfo{person}{Will Hamilton}, \bibinfo{person}{Zhitao
  Ying}, {and} \bibinfo{person}{Jure Leskovec}.}
  \bibinfo{year}{2017}\natexlab{}.
\newblock \showarticletitle{Inductive Representation Learning on Large Graphs}.
  In \bibinfo{booktitle}{\emph{Advances in Neural Information Processing
  Systems}}, \bibfield{editor}{\bibinfo{person}{I.~Guyon},
  \bibinfo{person}{U.~Von Luxburg}, \bibinfo{person}{S.~Bengio},
  \bibinfo{person}{H.~Wallach}, \bibinfo{person}{R.~Fergus},
  \bibinfo{person}{S.~Vishwanathan}, {and} \bibinfo{person}{R.~Garnett}}
  (Eds.), Vol.~\bibinfo{volume}{30}. \bibinfo{publisher}{Curran Associates,
  Inc.}
\newblock
\urldef\tempurl%
\url{https://proceedings.neurips.cc/paper_files/paper/2017/file/5dd9db5e033da9c6fb5ba83c7a7ebea9-Paper.pdf}
\showURL{%
\tempurl}


\bibitem[Hay et~al\mbox{.}(2009a)]%
        {hay2009accurate}
\bibfield{author}{\bibinfo{person}{Michael Hay}, \bibinfo{person}{Chao Li},
  \bibinfo{person}{Gerome Miklau}, {and} \bibinfo{person}{David Jensen}.}
  \bibinfo{year}{2009}\natexlab{a}.
\newblock \showarticletitle{Accurate estimation of the degree distribution of
  private networks}. In \bibinfo{booktitle}{\emph{2009 Ninth IEEE International
  Conference on Data Mining}}. IEEE, \bibinfo{pages}{169--178}.
\newblock


\bibitem[Hay et~al\mbox{.}(2009b)]%
        {Hay_degree_dist}
\bibfield{author}{\bibinfo{person}{Michael Hay}, \bibinfo{person}{Chao Li},
  \bibinfo{person}{Gerome Miklau}, {and} \bibinfo{person}{David~D. Jensen}.}
  \bibinfo{year}{2009}\natexlab{b}.
\newblock \showarticletitle{Accurate Estimation of the Degree Distribution of
  Private Networks}. In \bibinfo{booktitle}{\emph{{ICDM} 2009, The Ninth {IEEE}
  International Conference on Data Mining, Miami, Florida, USA, 6-9 December
  2009}}, \bibfield{editor}{\bibinfo{person}{Wei Wang}, \bibinfo{person}{Hillol
  Kargupta}, \bibinfo{person}{Sanjay Ranka}, \bibinfo{person}{Philip~S. Yu},
  {and} \bibinfo{person}{Xindong Wu}} (Eds.). \bibinfo{publisher}{{IEEE}
  Computer Society}, \bibinfo{pages}{169--178}.
\newblock
\urldef\tempurl%
\url{https://doi.org/10.1109/ICDM.2009.11}
\showDOI{\tempurl}


\bibitem[He et~al\mbox{.}(2021)]%
        {he2021stealing}
\bibfield{author}{\bibinfo{person}{Xinlei He}, \bibinfo{person}{Jinyuan Jia},
  \bibinfo{person}{Michael Backes}, \bibinfo{person}{Neil~Zhenqiang Gong},
  {and} \bibinfo{person}{Yang Zhang}.} \bibinfo{year}{2021}\natexlab{}.
\newblock \showarticletitle{Stealing Links from Graph Neural Networks.}. In
  \bibinfo{booktitle}{\emph{USENIX Security Symposium}}.
  \bibinfo{pages}{2669--2686}.
\newblock


\bibitem[Karwa and Slavkovic(2012)]%
        {Karwa_subgraph_count}
\bibfield{author}{\bibinfo{person}{Vishesh Karwa} {and}
  \bibinfo{person}{Aleksandra~B. Slavkovic}.} \bibinfo{year}{2012}\natexlab{}.
\newblock \showarticletitle{Differentially Private Graphical Degree Sequences
  and Synthetic Graphs}. In \bibinfo{booktitle}{\emph{Privacy in Statistical
  Databases - {UNESCO} Chair in Data Privacy, International Conference, {PSD}
  2012, Palermo, Italy, September 26-28, 2012. Proceedings}}
  \emph{(\bibinfo{series}{Lecture Notes in Computer Science},
  Vol.~\bibinfo{volume}{7556})}, \bibfield{editor}{\bibinfo{person}{Josep
  Domingo{-}Ferrer} {and} \bibinfo{person}{Ilenia Tinnirello}} (Eds.).
  \bibinfo{publisher}{Springer}, \bibinfo{pages}{273--285}.
\newblock
\urldef\tempurl%
\url{https://doi.org/10.1007/978-3-642-33627-0\_21}
\showDOI{\tempurl}


\bibitem[Kipf and Welling(2017a)]%
        {gcn_first}
\bibfield{author}{\bibinfo{person}{Thomas~N. Kipf} {and} \bibinfo{person}{Max
  Welling}.} \bibinfo{year}{2017}\natexlab{a}.
\newblock \showarticletitle{Semi-Supervised Classification with Graph
  Convolutional Networks}. In \bibinfo{booktitle}{\emph{5th International
  Conference on Learning Representations, {ICLR} 2017, Toulon, France, April
  24-26, 2017, Conference Track Proceedings}}.
  \bibinfo{publisher}{OpenReview.net}.
\newblock
\urldef\tempurl%
\url{https://openreview.net/forum?id=SJU4ayYgl}
\showURL{%
\tempurl}


\bibitem[Kipf and Welling(2017b)]%
        {gnn2}
\bibfield{author}{\bibinfo{person}{Thomas~N. Kipf} {and} \bibinfo{person}{Max
  Welling}.} \bibinfo{year}{2017}\natexlab{b}.
\newblock \showarticletitle{{Semi-Supervised Classification with Graph
  Convolutional Networks}}. In \bibinfo{booktitle}{\emph{Proceedings of the 5th
  International Conference on Learning Representations}} (Palais des
  Congr{\`e}s Neptune, Toulon, France) \emph{(\bibinfo{series}{ICLR '17})}.
\newblock
\urldef\tempurl%
\url{https://openreview.net/forum?id=SJU4ayYgl}
\showURL{%
\tempurl}


\bibitem[Kossinets and Watts(2006)]%
        {kossinets2006empirical}
\bibfield{author}{\bibinfo{person}{Gueorgi Kossinets} {and}
  \bibinfo{person}{Duncan~J Watts}.} \bibinfo{year}{2006}\natexlab{}.
\newblock \showarticletitle{Empirical analysis of an evolving social network}.
\newblock \bibinfo{journal}{\emph{science}} \bibinfo{volume}{311},
  \bibinfo{number}{5757} (\bibinfo{year}{2006}), \bibinfo{pages}{88--90}.
\newblock


\bibitem[Mu et~al\mbox{.}(2021)]%
        {mu2021hard_adversarial}
\bibfield{author}{\bibinfo{person}{Jiaming Mu}, \bibinfo{person}{Binghui Wang},
  \bibinfo{person}{Qi Li}, \bibinfo{person}{Kun Sun}, \bibinfo{person}{Mingwei
  Xu}, {and} \bibinfo{person}{Zhuotao Liu}.} \bibinfo{year}{2021}\natexlab{}.
\newblock \showarticletitle{A hard label black-box adversarial attack against
  graph neural networks}. In \bibinfo{booktitle}{\emph{Proceedings of the 2021
  ACM SIGSAC Conference on Computer and Communications Security}}.
  \bibinfo{pages}{108--125}.
\newblock


\bibitem[Mueller et~al\mbox{.}(2022)]%
        {sok_DP}
\bibfield{author}{\bibinfo{person}{Tamara~T. Mueller}, \bibinfo{person}{Dmitrii
  Usynin}, \bibinfo{person}{Johannes~C. Paetzold}, \bibinfo{person}{Daniel
  Rueckert}, {and} \bibinfo{person}{Georgios Kaissis}.}
  \bibinfo{year}{2022}\natexlab{}.
\newblock \showarticletitle{SoK: Differential Privacy on Graph-Structured
  Data}.
\newblock \bibinfo{journal}{\emph{CoRR}}  \bibinfo{volume}{abs/2203.09205}
  (\bibinfo{year}{2022}).
\newblock
\urldef\tempurl%
\url{https://doi.org/10.48550/arXiv.2203.09205}
\showDOI{\tempurl}
\showeprint[arXiv]{2203.09205}


\bibitem[Nguyen et~al\mbox{.}(2015)]%
        {Hiep_adj_release}
\bibfield{author}{\bibinfo{person}{Hiep~H. Nguyen}, \bibinfo{person}{Abdessamad
  Imine}, {and} \bibinfo{person}{Micha{\"{e}}l Rusinowitch}.}
  \bibinfo{year}{2015}\natexlab{}.
\newblock \showarticletitle{Differentially Private Publication of Social Graphs
  at Linear Cost}. In \bibinfo{booktitle}{\emph{Proceedings of the 2015
  {IEEE/ACM} International Conference on Advances in Social Networks Analysis
  and Mining, {ASONAM} 2015, Paris, France, August 25 - 28, 2015}},
  \bibfield{editor}{\bibinfo{person}{Jian Pei}, \bibinfo{person}{Fabrizio
  Silvestri}, {and} \bibinfo{person}{Jie Tang}} (Eds.).
  \bibinfo{publisher}{{ACM}}, \bibinfo{pages}{596--599}.
\newblock
\urldef\tempurl%
\url{https://doi.org/10.1145/2808797.2809385}
\showDOI{\tempurl}


\bibitem[Olatunji et~al\mbox{.}(2021)]%
        {olatunji2021MIA}
\bibfield{author}{\bibinfo{person}{Iyiola~E Olatunji},
  \bibinfo{person}{Wolfgang Nejdl}, {and} \bibinfo{person}{Megha Khosla}.}
  \bibinfo{year}{2021}\natexlab{}.
\newblock \showarticletitle{Membership inference attack on graph neural
  networks}. In \bibinfo{booktitle}{\emph{2021 Third IEEE International
  Conference on Trust, Privacy and Security in Intelligent Systems and
  Applications (TPS-ISA)}}. IEEE, \bibinfo{pages}{11--20}.
\newblock


\bibitem[Rozemberczki and Sarkar(2021)]%
        {rozemberczki2021twitch}
\bibfield{author}{\bibinfo{person}{Benedek Rozemberczki} {and}
  \bibinfo{person}{Rik Sarkar}.} \bibinfo{year}{2021}\natexlab{}.
\newblock \showarticletitle{Twitch gamers: a dataset for evaluating proximity
  preserving and structural role-based node embeddings}.
\newblock \bibinfo{journal}{\emph{arXiv preprint arXiv:2101.03091}}
  (\bibinfo{year}{2021}).
\newblock


\bibitem[Sajadmanesh et~al\mbox{.}(2022)]%
        {DP_GAP}
\bibfield{author}{\bibinfo{person}{Sina Sajadmanesh},
  \bibinfo{person}{Ali~Shahin Shamsabadi}, \bibinfo{person}{Aur{\'{e}}lien
  Bellet}, {and} \bibinfo{person}{Daniel Gatica{-}Perez}.}
  \bibinfo{year}{2022}\natexlab{}.
\newblock \showarticletitle{{GAP:} Differentially Private Graph Neural Networks
  with Aggregation Perturbation}.
\newblock \bibinfo{journal}{\emph{CoRR}}  \bibinfo{volume}{abs/2203.00949}
  (\bibinfo{year}{2022}).
\newblock
\urldef\tempurl%
\url{https://doi.org/10.48550/arXiv.2203.00949}
\showDOI{\tempurl}
\showeprint[arXiv]{2203.00949}


\bibitem[Scarselli et~al\mbox{.}(2009)]%
        {gnn_first}
\bibfield{author}{\bibinfo{person}{Franco Scarselli}, \bibinfo{person}{Marco
  Gori}, \bibinfo{person}{Ah~Chung Tsoi}, \bibinfo{person}{Markus
  Hagenbuchner}, {and} \bibinfo{person}{Gabriele Monfardini}.}
  \bibinfo{year}{2009}\natexlab{}.
\newblock \showarticletitle{The Graph Neural Network Model}.
\newblock \bibinfo{journal}{\emph{IEEE Transactions on Neural Networks}}
  \bibinfo{volume}{20}, \bibinfo{number}{1} (\bibinfo{date}{Jan}
  \bibinfo{year}{2009}), \bibinfo{pages}{61--80}.
\newblock
\showISSN{1941-0093}
\urldef\tempurl%
\url{https://doi.org/10.1109/TNN.2008.2005605}
\showDOI{\tempurl}


\bibitem[Sen et~al\mbox{.}(2008)]%
        {citation_datasets}
\bibfield{author}{\bibinfo{person}{Prithviraj Sen}, \bibinfo{person}{Galileo
  Namata}, \bibinfo{person}{Mustafa Bilgic}, \bibinfo{person}{Lise Getoor},
  \bibinfo{person}{Brian Galligher}, {and} \bibinfo{person}{Tina Eliassi-Rad}.}
  \bibinfo{year}{2008}\natexlab{}.
\newblock \showarticletitle{Collective Classification in Network Data}.
\newblock \bibinfo{journal}{\emph{AI Magazine}} \bibinfo{volume}{29},
  \bibinfo{number}{3} (\bibinfo{date}{Sep.} \bibinfo{year}{2008}),
  \bibinfo{pages}{93}.
\newblock
\urldef\tempurl%
\url{https://doi.org/10.1609/aimag.v29i3.2157}
\showDOI{\tempurl}


\bibitem[Sun et~al\mbox{.}(2018)]%
        {sun2018adversarial_survey}
\bibfield{author}{\bibinfo{person}{Lichao Sun}, \bibinfo{person}{Yingtong Dou},
  \bibinfo{person}{Carl Yang}, \bibinfo{person}{Ji Wang},
  \bibinfo{person}{Yixin Liu}, \bibinfo{person}{Philip~S Yu},
  \bibinfo{person}{Lifang He}, {and} \bibinfo{person}{Bo Li}.}
  \bibinfo{year}{2018}\natexlab{}.
\newblock \showarticletitle{Adversarial attack and defense on graph data: A
  survey}.
\newblock \bibinfo{journal}{\emph{arXiv preprint arXiv:1812.10528}}
  (\bibinfo{year}{2018}).
\newblock


\bibitem[Velickovic et~al\mbox{.}(2018)]%
        {Petar_GTA_node_class}
\bibfield{author}{\bibinfo{person}{Petar Velickovic}, \bibinfo{person}{Guillem
  Cucurull}, \bibinfo{person}{Arantxa Casanova}, \bibinfo{person}{Adriana
  Romero}, \bibinfo{person}{Pietro Li{\`{o}}}, {and} \bibinfo{person}{Yoshua
  Bengio}.} \bibinfo{year}{2018}\natexlab{}.
\newblock \showarticletitle{Graph Attention Networks}. In
  \bibinfo{booktitle}{\emph{6th International Conference on Learning
  Representations, {ICLR} 2018, Vancouver, BC, Canada, April 30 - May 3, 2018,
  Conference Track Proceedings}}. \bibinfo{publisher}{OpenReview.net}.
\newblock
\urldef\tempurl%
\url{https://openreview.net/forum?id=rJXMpikCZ}
\showURL{%
\tempurl}


\bibitem[Wang et~al\mbox{.}(2022)]%
        {Wang_graph_class}
\bibfield{author}{\bibinfo{person}{Zhaohui Wang}, \bibinfo{person}{Qi Cao},
  \bibinfo{person}{Huawei Shen}, \bibinfo{person}{Bingbing Xu},
  \bibinfo{person}{Muhan Zhang}, {and} \bibinfo{person}{Xueqi Cheng}.}
  \bibinfo{year}{2022}\natexlab{}.
\newblock \showarticletitle{Towards Efficient and Expressive GNNs for Graph
  Classification via Subgraph-Aware Weisfeiler-Lehman}. In
  \bibinfo{booktitle}{\emph{Learning on Graphs Conference, LoG 2022, 9-12
  December 2022, Virtual Event}} \emph{(\bibinfo{series}{Proceedings of Machine
  Learning Research}, Vol.~\bibinfo{volume}{198})},
  \bibfield{editor}{\bibinfo{person}{Bastian Rieck} {and}
  \bibinfo{person}{Razvan Pascanu}} (Eds.). \bibinfo{publisher}{{PMLR}},
  \bibinfo{pages}{17}.
\newblock
\urldef\tempurl%
\url{https://proceedings.mlr.press/v198/wang22b.html}
\showURL{%
\tempurl}


\bibitem[Wu et~al\mbox{.}(2021a)]%
        {MIA_GNN}
\bibfield{author}{\bibinfo{person}{Bang Wu}, \bibinfo{person}{Xiangwen Yang},
  \bibinfo{person}{Shirui Pan}, {and} \bibinfo{person}{Xingliang Yuan}.}
  \bibinfo{year}{2021}\natexlab{a}.
\newblock \showarticletitle{Adapting Membership Inference Attacks to {GNN} for
  Graph Classification: Approaches and Implications}. In
  \bibinfo{booktitle}{\emph{{IEEE} International Conference on Data Mining,
  {ICDM} 2021, Auckland, New Zealand, December 7-10, 2021}},
  \bibfield{editor}{\bibinfo{person}{James Bailey}, \bibinfo{person}{Pauli
  Miettinen}, \bibinfo{person}{Yun~Sing Koh}, \bibinfo{person}{Dacheng Tao},
  {and} \bibinfo{person}{Xindong Wu}} (Eds.). \bibinfo{publisher}{{IEEE}},
  \bibinfo{pages}{1421--1426}.
\newblock
\urldef\tempurl%
\url{https://doi.org/10.1109/ICDM51629.2021.00182}
\showDOI{\tempurl}


\bibitem[Wu et~al\mbox{.}(2021b)]%
        {wu2021MIA}
\bibfield{author}{\bibinfo{person}{Bang Wu}, \bibinfo{person}{Xiangwen Yang},
  \bibinfo{person}{Shirui Pan}, {and} \bibinfo{person}{Xingliang Yuan}.}
  \bibinfo{year}{2021}\natexlab{b}.
\newblock \showarticletitle{Adapting membership inference attacks to GNN for
  graph classification: approaches and implications}. In
  \bibinfo{booktitle}{\emph{2021 IEEE International Conference on Data Mining
  (ICDM)}}. IEEE, \bibinfo{pages}{1421--1426}.
\newblock


\bibitem[Wu et~al\mbox{.}(2022)]%
        {linkteller}
\bibfield{author}{\bibinfo{person}{Fan Wu}, \bibinfo{person}{Yunhui Long},
  \bibinfo{person}{Ce Zhang}, {and} \bibinfo{person}{Bo Li}.}
  \bibinfo{year}{2022}\natexlab{}.
\newblock \showarticletitle{Linkteller: Recovering private edges from graph
  neural networks via influence analysis}. In \bibinfo{booktitle}{\emph{2022
  IEEE Symposium on Security and Privacy (SP)}}. IEEE,
  \bibinfo{pages}{2005--2024}.
\newblock


\bibitem[Xu et~al\mbox{.}(2019)]%
        {Xu_graph_class}
\bibfield{author}{\bibinfo{person}{Keyulu Xu}, \bibinfo{person}{Weihua Hu},
  \bibinfo{person}{Jure Leskovec}, {and} \bibinfo{person}{Stefanie Jegelka}.}
  \bibinfo{year}{2019}\natexlab{}.
\newblock \showarticletitle{How Powerful are Graph Neural Networks?}. In
  \bibinfo{booktitle}{\emph{7th International Conference on Learning
  Representations, {ICLR} 2019, New Orleans, LA, USA, May 6-9, 2019}}.
  \bibinfo{publisher}{OpenReview.net}.
\newblock
\urldef\tempurl%
\url{https://openreview.net/forum?id=ryGs6iA5Km}
\showURL{%
\tempurl}


\bibitem[Ye et~al\mbox{.}(2022)]%
        {MIA_Jiayuan}
\bibfield{author}{\bibinfo{person}{Jiayuan Ye}, \bibinfo{person}{Aadyaa Maddi},
  \bibinfo{person}{Sasi~Kumar Murakonda}, \bibinfo{person}{Vincent
  Bindschaedler}, {and} \bibinfo{person}{Reza Shokri}.}
  \bibinfo{year}{2022}\natexlab{}.
\newblock \showarticletitle{Enhanced Membership Inference Attacks against
  Machine Learning Models}. In \bibinfo{booktitle}{\emph{Proceedings of the
  2022 ACM SIGSAC Conference on Computer and Communications Security}} (Los
  Angeles, CA, USA) \emph{(\bibinfo{series}{CCS '22})}.
  \bibinfo{publisher}{Association for Computing Machinery},
  \bibinfo{address}{New York, NY, USA}, \bibinfo{pages}{3093–3106}.
\newblock
\showISBNx{9781450394505}
\urldef\tempurl%
\url{https://doi.org/10.1145/3548606.3560675}
\showDOI{\tempurl}


\bibitem[Zeng et~al\mbox{.}(2020)]%
        {ZengZSKP20_Flickr}
\bibfield{author}{\bibinfo{person}{Hanqing Zeng}, \bibinfo{person}{Hongkuan
  Zhou}, \bibinfo{person}{Ajitesh Srivastava}, \bibinfo{person}{Rajgopal
  Kannan}, {and} \bibinfo{person}{Viktor~K. Prasanna}.}
  \bibinfo{year}{2020}\natexlab{}.
\newblock \showarticletitle{GraphSAINT: Graph Sampling Based Inductive Learning
  Method}. In \bibinfo{booktitle}{\emph{8th International Conference on
  Learning Representations, {ICLR} 2020, Addis Ababa, Ethiopia, April 26-30,
  2020}}. \bibinfo{publisher}{OpenReview.net}.
\newblock
\urldef\tempurl%
\url{https://openreview.net/forum?id=BJe8pkHFwS}
\showURL{%
\tempurl}


\bibitem[Zhang and Chen(2018)]%
        {Zhang_link_pred}
\bibfield{author}{\bibinfo{person}{Muhan Zhang} {and} \bibinfo{person}{Yixin
  Chen}.} \bibinfo{year}{2018}\natexlab{}.
\newblock \showarticletitle{Link prediction based on graph neural networks}.
\newblock \bibinfo{journal}{\emph{Advances in neural information processing
  systems}}  \bibinfo{volume}{31} (\bibinfo{year}{2018}).
\newblock


\bibitem[Zhang et~al\mbox{.}(2020)]%
        {zhang2020_adversarial}
\bibfield{author}{\bibinfo{person}{Mengmei Zhang}, \bibinfo{person}{Linmei Hu},
  \bibinfo{person}{Chuan Shi}, {and} \bibinfo{person}{Xiao Wang}.}
  \bibinfo{year}{2020}\natexlab{}.
\newblock \showarticletitle{Adversarial label-flipping attack and defense for
  graph neural networks}. In \bibinfo{booktitle}{\emph{2020 IEEE International
  Conference on Data Mining (ICDM)}}. IEEE, \bibinfo{pages}{791--800}.
\newblock


\bibitem[Zhu et~al\mbox{.}(2022)]%
        {postprocessing}
\bibfield{author}{\bibinfo{person}{Keyu Zhu}, \bibinfo{person}{Ferdinando
  Fioretto}, {and} \bibinfo{person}{Pascal Van~Hentenryck}.}
  \bibinfo{year}{2022}\natexlab{}.
\newblock \showarticletitle{Post-processing of Differentially Private Data: A
  Fairness Perspective}. In \bibinfo{booktitle}{\emph{Proceedings of the
  Thirty-First International Joint Conference on Artificial Intelligence,
  {IJCAI-22}}}, \bibfield{editor}{\bibinfo{person}{Lud~De Raedt}} (Ed.).
  \bibinfo{publisher}{International Joint Conferences on Artificial
  Intelligence Organization}, \bibinfo{pages}{4029--4035}.
\newblock
\urldef\tempurl%
\url{https://doi.org/10.24963/ijcai.2022/559}
\showDOI{\tempurl}
\newblock
\shownote{Main Track}.


\end{thebibliography}

\end{document}